\documentclass[12pt]{iopart}
\usepackage{iopams} 
\usepackage{bm}
\usepackage{graphicx}
\usepackage{setstack}
\usepackage{color}
\usepackage{bm}
\usepackage{latexsym}
\usepackage{iopams}
\usepackage{delarray}
\usepackage{amssymb}

\begin{document}

\title{Advanced Interacting Sequential Monte Carlo Sampling for Inverse Scattering}
\author{F Giraud$^{1,2}$, P Minvielle$^1$ and P Del Moral$^{2}$}
\address{$^1$ CEA-CESTA, 33114 Le Barp, France}
\address{$^2$ INRIA Bordeaux Sud-Ouest, Domaine Universitaire, 351, cours de la Lib\'{e}ration, 33405 Talence Cedex, France}
\ead{francois.giraud@ens-cachan.org}
\ead{pierre.minvielle@cea.fr}


\begin{abstract}
The following electromagnetism (EM) inverse problem is addressed. It consists in estimating local radioelectric properties of materials recovering an object from global EM scattering measurements, at various incidences and wave frequencies. This large scale ill-posed inverse problem is explored by an intensive exploitation of an efficient 2D Maxwell solver, distributed on high performance computing machines. Applied to a large training data set, a statistical analysis reduces the problem to a simpler probabilistic metamodel, on which Bayesian inference can be performed.  Considering the radioelectric properties  as a hidden dynamic stochastic process, that evolves in function of the frequency, it is shown how advanced Markov Chain Monte Carlo methods, called Sequential Monte Carlo (SMC) or interacting particles, can take benefit of the structure and provide local EM property estimates.
\end{abstract}


\section{Introduction}
\paragraph*{}
Inverse scattering is a topic of major importance; it encompasses various applications \cite{hopcraft1992introduction,colton2003linear} in acoustics, optics and electromagnetism, e.g. medical imaging, tomography, ionospheric sounding or SAR (Synthetic Aperture Radar). In electromagnetism (EM), the direct scattering problem is the determination of the scattered field, due to the scattering of an incident wave in the presence of inhomogeneities, when the geometrical and physical properties of the scatterer are known. Conversely, inverse scattering is defined as "inferring information on the inhomogeneity from knowledge of the far-field pattern..."  \cite{colton2003linear}; it is an inverse problem. In this paper, we focus on a specific, though worthwhile, EM inverse scattering issue. The aim is to estimate the electromagnetic properties of materials from global microwave scattering measurements. Related applications can be located at the crossroads of non-destructive testing, quality control and material measurement. Many EM material characterization techniques have been developed in  the domain of agricultural and food materials, radar absorbers \cite{knott2004radar}, etc. Most of these techniques, from  the transmission lines to the admittance tunnel method, require small-scale material test samples. For instance, transmission lines enclosed samples inside the conductors of a transmission-line sample holder.  Although the EM properties (i.e. permeability  and permittivity) can be measured, they can differ significantly from the final product's ones, when the materials are assembled and placed on the full-scaled object or system \cite{knott2004radar}. The so-called free-space RCS (Radar Cross Section: scalar that quantifies reflectivity) methods \cite{knott2004radar} can overcome this pitfall by measuring the monostatic reflectivity of a large planar sample. The sample is then located inside an anechoic chamber, in the far field of the transmitting and receiving antennas. The reflectivity is measured at various arrival angles of the incident wave. Besides, let mention the classic bistatic alternative in near field, known as  the NRL arch method \cite{knott2004radar}.  In this paper, we focus on the following challenging inverse scattering problem:  the control and evaluation of EM properties of a full-scaled objet or mock-up from the global reflectivity measurements in a free-space RCS device. Deviations of microwave properties, such as permeability and permittivity, are to be determined along the object.

\paragraph*{}
Nearly 50 years ago, a closely related issue was formerly outlined in \cite{abbato1965dielectric}. Least-square optimization was applied to determine the dielectric constants that made the analytically computed RCS fit with measurements. This issue reemerged in a slightly different way  in \cite{coen1981inverse} ; both complex permittivity and permeability of a lossless plane stratified medium were evaluated. More recently,  \cite{garnero1991microwave} considers the reconstruction in microwave tomography of the dielectric properties of a strongly inhomogeneous object  by a  stochastic global optimization algorithm, based on simulated annealing. Similarly, \cite{caorsi1990two} develops a pseudoinversion algorithm  for 2D imaging, with the aim  locating and estimating the dielectric permittivities of unknown inhomogeneous dielectric cylindrical objects. On the whole, inverse scattering is known to be an ill-posed inverse problem. Like image reconstruction and many other imaging inverse problems \cite{ribes2008linear,demoment1989image}, it necessitates at some step  a regularization procedure: it tends to eliminate the artificial oscillations resulting from to the problem ill-posedness.  According to \cite{colton2003linear}, the procedures can be partitioned into the next two families: the non-linear optimization schemes and weak scattering linearization approximation methods, such as physical optics and Born approximation. Besides, let mention the efficient linear sampling method in 3D shape reconstruction of obstacles due to local inhomogeneities \cite{collino2003numerical,colton2003linear}.  

\paragraph*{}
In this paper, a global statistical approach is developed to address the "free space RCS" inverse scattering and solve the large scale ill-posed inverse problem. In some way, the approach can be considered to be part of the two aforementioned procedure families. It involves an approximation method. Intensive Maxwell solver computations, distributed on high performance computing (HPC) machines, results in a surrogate likelihood model. It is the starting point of a complete statistical dynamic model framework that leads to an efficient inference scheme, close to optimization. It stems from statistical signal processing and advanced Monte Carlo sampling (e.g. Markov Chain Monte Carlo). Bayesian inference is performed by a sequential Monte Carlo (SMC) stochastic algorithm.  These algorithms are called "interacting particles" \cite{DMDJ} or particle filtering in adaptive filtering and sequential estimation. They are used to provide, in addition to microwave properties estimates of materials, the very significant information of the associated uncertainties. From the seminal  work of Geman and Geman \cite{geman1984stochastic}, stochastic methods have been commonly used in inverse scattering and, more generally, in image inverse problems: simulated annealing for image reconstruction \cite{robini1999simulated}, expectation-maximization algorithm for radar imaging \cite{lanterman2000statistical}, etc.  In microwave imaging, \cite{pastorino2007stochastic} points out genetic algorithms and stochastic heuristics, such as differential evolution methods, memetic algorithms, particle swarm optimizations, ant colonies, etc. In short, many attempts have been made in electromagnetism to apply stochastic methods to tricky inverse problems or non-convex optimization (such as multilayered radar absorbing coatings  \cite{michielssen1993design,chambers1996optimised}). Though powerful, stochastic inverse methods  often come up against high-dimensional curse. In the approach, it is taken advantage of the problem structure to achieve a Rao-Blackwellisation strategy \cite{Liu,DMDJ} of Monte Carlo variance reduction and design a powerful stochastic inversion method that overcomes the high-dimensional obstacle. 

\paragraph*{}
This paper is organized as follows. In section \ref{SecPb}, free-space RCS material measurements is introduced and the inverse scattering problem is developed. Next,  section  \ref{SecMod} describes the probabilistic modeling , from the surrogate likelihood model to the overall statistical dynamic modeling framework and, at its core, a hidden Markov model (HMM). The inversion  Rao-Blackwellised stochastic algorithm is developed in section  \ref{sec-traitement}. It is evaluated in section  \ref{sec-applications} where its statistical performance is assessed. 


\section{The inverse scattering problem} \label{SecPb}

\subsection{Electromagnetic scattering measurement}  \label{SecPbEMmeas}
\begin{figure}[ht!] 
\centerline{\includegraphics[width=0.4\columnwidth]{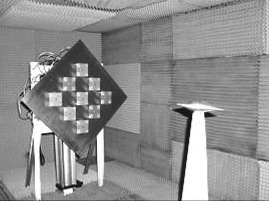}} \caption{RCS measurement inside an anechoic chamber} \label{RCSmeas} 
\end{figure}

\paragraph*{}
EM scattering measurements have been achieved ever since radar invention \cite{knott2004radar}. Briefly speaking, EM scattering is the standard phenomenon that occurs when an object is exposed to an EM wave and disperses incident energy in all directions (scattering is this spatial distribution of energy). Some energy is scattered back to the source of the wave. It constitutes the radar echo of the object, the intensity of which results from the radar cross section (RCS) of the object. More precisely, RCS is defined by:
\begin{equation}
\sigma_s=\lim\limits_{R \to +\infty} 4\pi R^2 \frac{\left| \mathbf{E}_\mathrm{scat} \right|^2}{\left| \mathbf{E}_\mathrm{inc} \right|^2} 
\label{defRCS} \end{equation}
It quantifies the scattering power of an object, i.e.  the ratio between the scattered power density $\mathbf{E}_\mathrm{scat}$ at the receiver and  the  power density of the incident wave at the target (with $R$  the radar-object range). It depends on the wave polarization and frequency. The  $4\pi R^2$ term takes into account the radiated spherical wave. Implicitly,  (\ref{defRCS}) requires that the incident wave is planar  ($R \to +\infty$). Practically, it is possible to measure the RCS at limited ranges with a sufficient accuracy. It is usually achieved in indoor RCS test chambers, also called anechoic chambers. There, interferences can be limited by microwave absorbing materials (see figure \ref{RCSmeas}). 

\begin{figure}[ht!] \centerline{ \includegraphics[width=0.6\columnwidth,clip=true,viewport=1cm 5cm 29cm 16cm,draft=false]{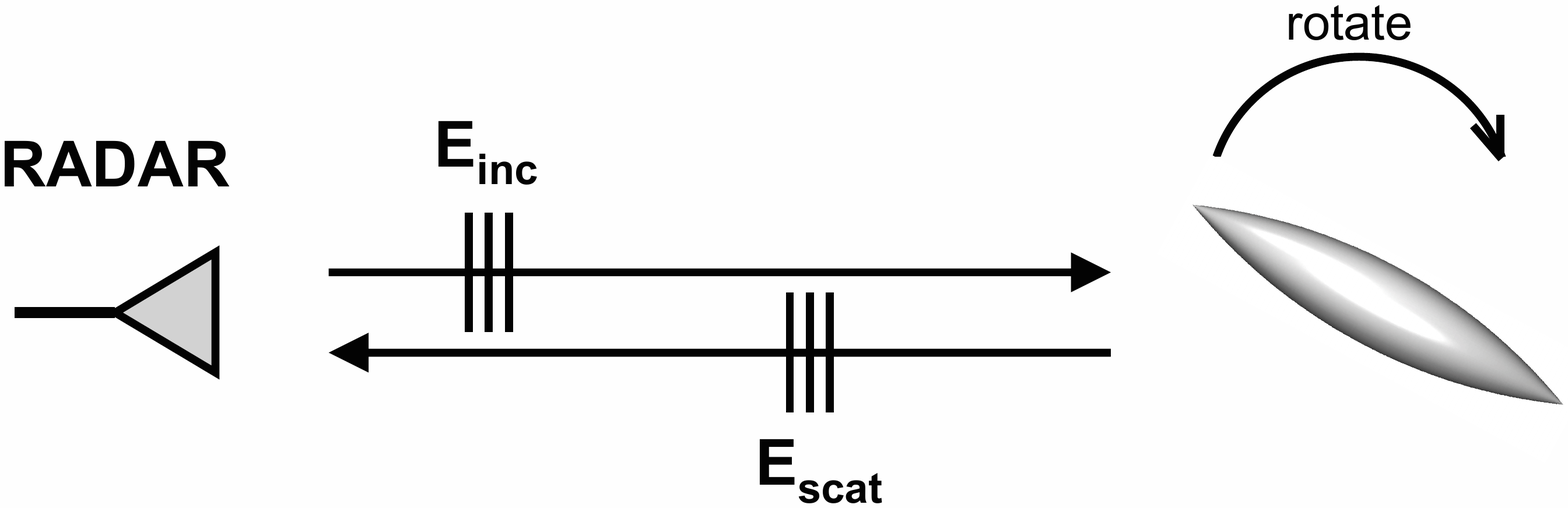}}  \caption{Monostatic scattering measurement principle} \label{RCSmeasprinc} \end{figure}

\paragraph*{}
In the article, we consider that an object or mock-up is illuminated by a radar, i.e. a single antenna or a more complex device (such as the antenna array of figure \ref{RCSmeas}) that fulfills to a certain extent directivity and far-field conditions \cite{minvielle2011sparse}. Herein the radar system is monostatic, which means that the transmitter and receiver are collocated. Its common principle is described in figure \ref{RCSmeasprinc}. Considering that the radar illuminates the object at a given incidence with a quasi-planar monochromatic continuous wave (CW) of frequency $f$ (incident electric field $\mathbf{E}_\mathrm{inc}$), the object backscatters a CW to the radar (scattered electric field $\mathbf{E}_\mathrm{scat}$) at the same frequency. With an appropriate  instrumentation system (radar, network analyzers, etc.) and a calibration process, it is possible to measure the complex scattering coefficient, which can be roughly defined by: $\mathcal{S} = \frac{\mathbf{E}_\mathrm{scat}}{\mathbf{E}_\mathrm{inc}}$. It sums up the EM scattering, indicating the wave change in amplitude and phase. $\mathcal{S}$ is closely linked to the RCS, with: $\sigma_s=|\mathcal{S}|^2$.  It is important to notice that the scattering coefficient quantifies a global characteristic of the whole object-EM wave interaction in specific conditions (incidence, frequency, etc.). It is possible to measure the scattering coefficient for different transmitted and received polarizations. 

\paragraph*{}
Let assume the following conventional RCS acquisition mode, widely used in Inverse Synthetic Aperture Radar (ISAR) imaging. It consists in measuring various complex scattering coefficients $\mathcal{S}$:
\begin{itemize}
\item[-] at different wave frequencies: $f\in\{f_1,f_2,\cdots,f_{K_f}\}$, for $K_f$ successive discrete frequencies. Basically, it consists in a series of transmitted narrow-band pulses, commonly known as SFCW (Stepped Frequency Continuous Wave) burst \cite{wehner1995artech}.
\item[-]   at different incidence angles: $\theta \in\{\theta_1,\theta_2,\cdots,\theta_{K_{\theta}}\}$, for $K_\theta$ different incidence angles (object rotation with a motorized rotating support). 
\item[-]  at different (transmitted and received) linear polarizations:  $\mathrm{pol}\in\{HH,VV\}$, meaning respectively, horizontally and vertically polarized both at microwave emission and reception.
\end{itemize}

Let call $\mathcal{M}$ the complete measurement, set of $2 \cdot K_f \cdot K_\theta$  elementary complex scattering coefficients: $\mathcal{M}=\{\mathcal{S}^{f,\theta,\mathrm{pol}}\}$, for $f\in\{f_1,\cdots,f_{K_f}\} \times \theta \in\{\theta_1,\cdots,\theta_{K_{\theta}}\}  \times \mathrm{pol}\in\{HH,VV\}$.

\subsection{Nondestructive testing}
\paragraph*{}
In this article, we are interested in an industrial control issue, that can be assimilated to nondestructive testing (NDT). Unlike usual EM material characterization techniques \cite{knott2004radar}, the point is to determine or check radioelectric properties (i.e. relative dielectric permittivity and magnetic permeability) of materials that are assembled and placed on the full-scaled object or system. Is it possible from the above complete measurement $\mathcal{M}$? Is it possible to extract some local information on the material properties along the object from the global scattering measurement information?

\begin{figure}[ht!]
 \centerline{ \includegraphics[width=0.5\columnwidth,clip=true,viewport=5cm 6cm 25cm 13cm,draft=false]{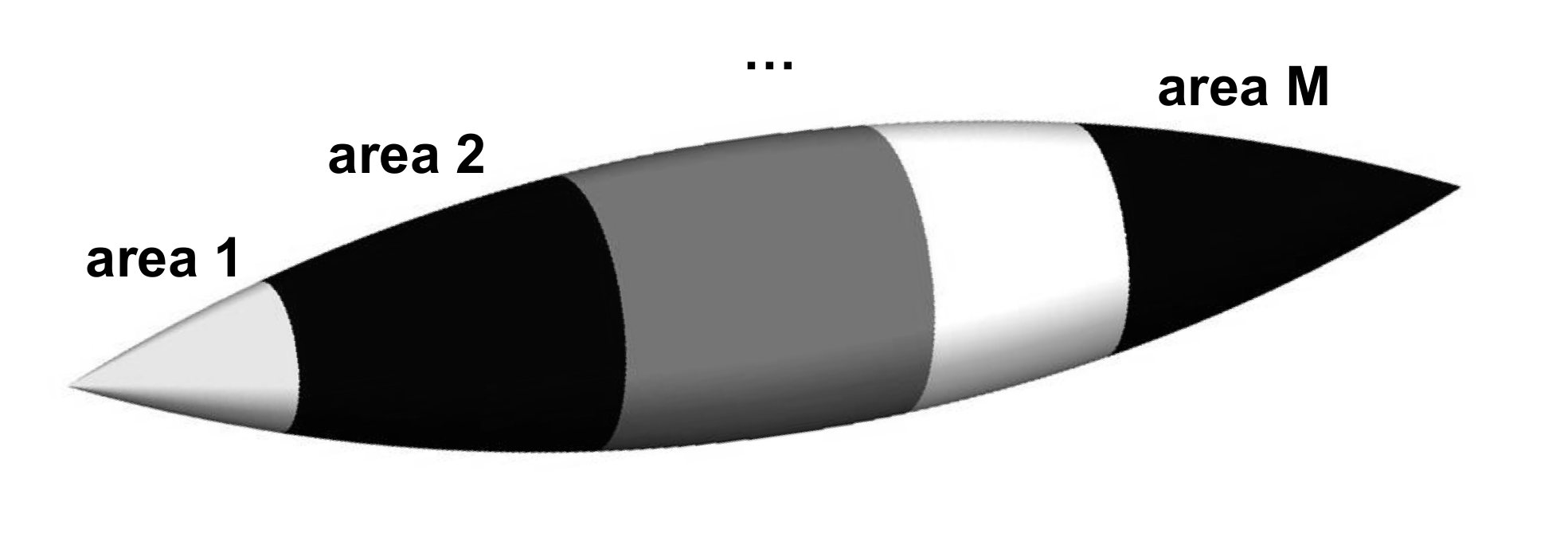}} 
\caption{The object coated by  $N_a$  material areas} \label{Ogive_Areas} 
\end{figure}

\begin{figure}[ht!]
 \centerline{ \includegraphics[width=0.5\columnwidth,clip=true,viewport=6cm 7cm 22cm 15cm,draft=false]{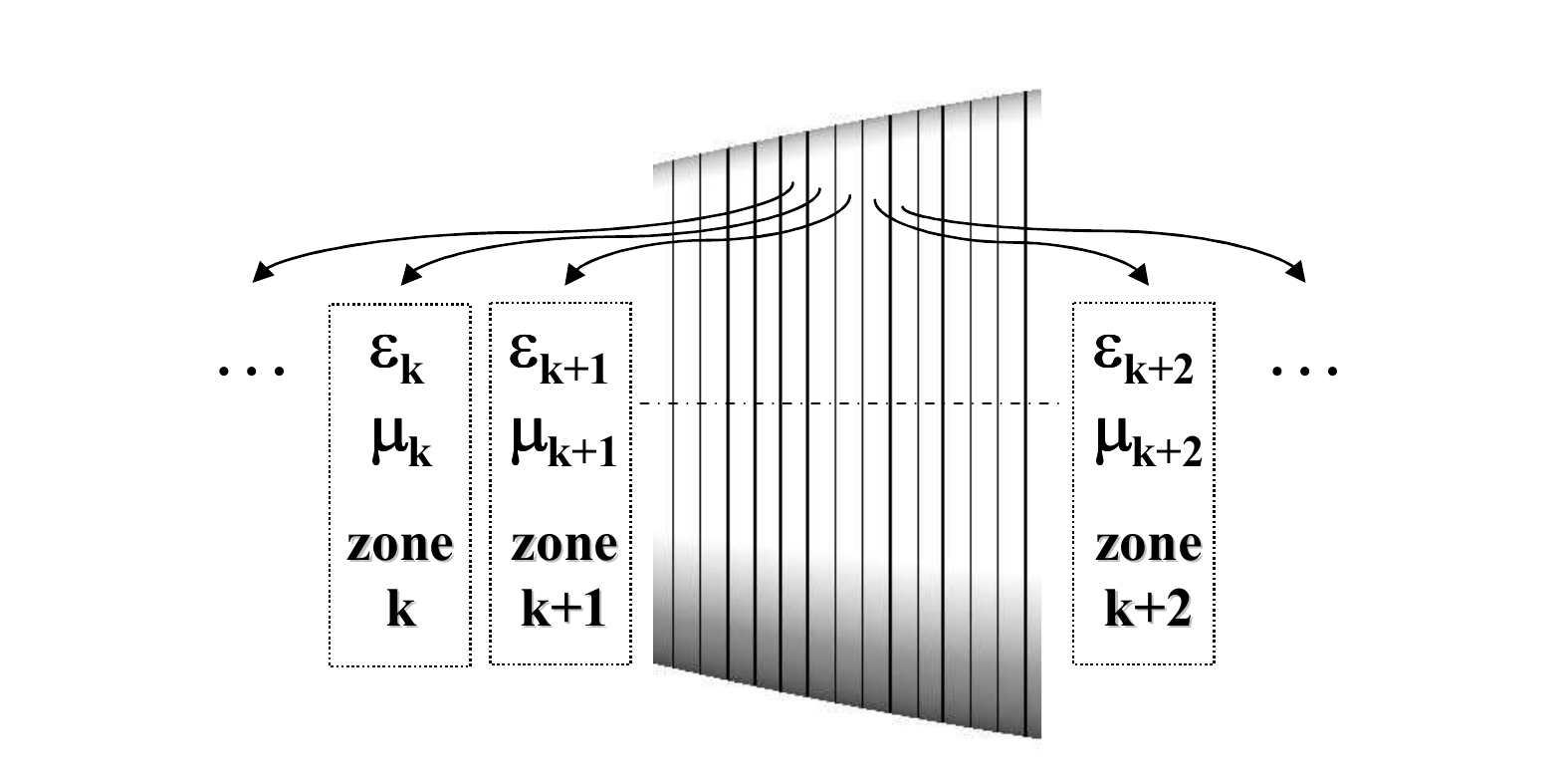}} 
\caption{Elementary mesh zones} \label{ElemAreas} 
\end{figure}

\paragraph*{}
In order to circumscribe the investigation, the article is restricted to a metallic axisymmetric object, which is is coated by $N_a$  material areas, each area corresponding to a rather homogeneous material, with its associated isotropic radioelectric properties weakly varying within the area. It is illustrated in figure \ref{Ogive_Areas}, with an ogival shape taken from the RCS benchmark \cite{woo1993programmer}.
Consequently, the aim is to determine from the global scattering measurement  $\mathcal{M}$ the unknown isotropic local EM properties $(\epsilon_1,\mu_1), (\epsilon_2,\mu_2), \cdots, (\epsilon_N,\mu_N)$ along the object,  where $N$ is the number of different elementary zones (cf. Figure \ref{ElemAreas}).

\subsection{An inverse problem for Maxwell's equations}
\paragraph*{}
Naturally, there is no direct model that is able to compute the radioelectric properties from global scattering information. On the contrary, the forward scattering model based on the resolution of Maxwell's equations can determine the scattering coefficients given the EM properties, the object geometry and acquisition conditions (i.e. wave frequency, incidence, etc.). It  lies in the resolution of Maxwell's equations, partial derivative equations that represent the electromagnetic scattering problem of an inhomogeneous obstacle.  It is performed by an efficient parallelized harmonic Maxwell  solver, an exact method that combines a volume finite element method and integral equation technique,  taking benefit from the axisymmetrical geometry of the shape \cite{stupfel1991combined}. Discretization is known to lead to problems of very large sizes, especially when the frequency is high. Furthermore, as it is shown further, the solver is  to be run many times for the inversion purpose. Hence, it necessitates high performance computing (HPC): a massive supercomputing system, with nearly  20,000 processors and a  performance higher than $1$ petaflops (million billion operations per second).

\begin{figure}[ht!] 
\includegraphics[width=0.8\columnwidth,clip=true,viewport=0cm 6.5cm 28cm 13cm,draft=false]{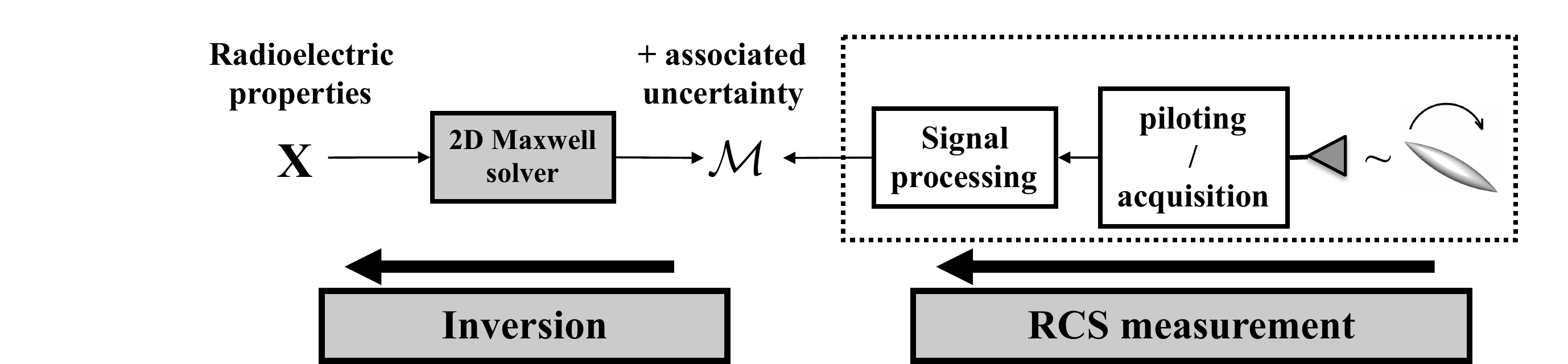} \caption{The inverse scattering problem} \label{PbInv}
\end{figure}

\paragraph*{}
Figure \ref{PbInv} sums up the entire inverse scattering problem. On one hand, the RCS measurement process, that includes acquisition, signal processing, calibration, etc., provides the complex scattering measurement $\mathcal{M}$, with uncertainties. On the other hand,  it would be useful to "row upstream" the Maxwell solver, in order to determine the unknown radioelectric properties, denoted by $\mathbf{x}$. Yet, even with recourse to HPC, there is no direct way to solve what turns out to be a high dimensional ill-posed inverse problem, like imaging inverse problems \cite{demoment1989image}. Next, we propose a global statistical inference approach, which is able to take into account prior information and achieve the required inversion. Like Tikhonov regularization, it tends to eliminate artificial oscillations due to the ill-posedness of the problem.  


\section{The statistical problem formulation} \label{SecMod}
The global statistical approach is introduced gradually, from its formulation at a given frequency $f_k$ to the whole stochastic model at the various frequencies   $f_1,f_2,\cdots,f_{K_f}$.

\subsection{The problem statement at  a single frequency $f_k$} 
Consider a given frequency $f_k$ of the SFCW burst. Let define the two main modeling components at $f_k$: the system state  $\mathbf{x}_k$, the observation $\mathbf{y}_k$ and the probabilistic link between them, i.e. the likelihood model $p(\mathbf{y}_k |\mathbf{x}_k )$.  To lighten the notations, they are  denoted respectively  $\mathbf{x}$, $\mathbf{y}$ and $p(\mathbf{y} |\mathbf{x})$ in this section. 

\subsubsection{System state}
$
\mathbf{x}=\left[  	\bm{\underline{\epsilon}}^{\prime}  \quad \bm{\underline{\epsilon}}^{\prime\prime}  \quad \bm{\underline{\mu}}^{\prime}  \quad \bm{\underline{\mu}}^{\prime\prime} \right]^T
$  includes the relative permittivity and permeability components of the $N$ elementary zones, where  $^{\prime}$ and $^{\prime\prime}$ denote respectively the real and imaginary parts \footnote{In other words, $\bm{\epsilon}=\bm{\epsilon}^{\prime}+j\bm{\epsilon}^{\prime\prime}$ and $\bm{\mu}=\bm{\mu}^{\prime}+j\bm{\mu}^{\prime\prime}$ (for time dependence convention $e^{j\omega t}$). } (at frequency $f_k$). The  four components can be developed  as: $\bm{\underline{\epsilon}}^{\prime}=\left[\epsilon^{\prime}_1  \cdots \epsilon^{\prime}_N \right]^T$, $
\bm{\underline{\epsilon}}^{\prime\prime}=\left[\epsilon^{\prime\prime}_1  \cdots \epsilon^{\prime\prime}_N \right]^T$, $
\bm{\underline{\mu}}^{\prime}=\left[\mu^{\prime}_1  \cdots \mu^{\prime}_N \right]^T$ and $
\bm{\underline{\mu}}^{\prime\prime}=\left[\mu^{\prime\prime}_1  \cdots \mu^{\prime\prime}_N \right]^T$. $\mathbf{x}$ is in a system space of  dimension $4N$; it includes all the unknown parameters that are to be estimated.

\subsubsection{Observation} $\mathbf{y}=\left[ \Re(\mathbf{\mathcal{S}_{HH}}) \quad \Im(\mathbf{\mathcal{S}_{HH}}) \quad \Re(\mathbf{\mathcal{S}_{VV}}) \quad \Im(\mathbf{\mathcal{S}_{VV}}) \right]^T$ contains the real ($\Re(\cdot)$) and imaginary  ($\Im(\cdot)$) parts of the complex scattering coefficients $\mathbf{\mathcal{S}_{HH}}$ et $\mathbf{\mathcal{S}_{VV}}$  measured at the $K_\theta$ angles $\theta_1, \cdots, \theta_{K_\theta}$ (at frequency $f_k$). The two complex terms $\mathbf{\mathcal{S}_{HH}}$ and $\mathbf{\mathcal{S}_{VV}}$ can be detailed: $\mathbf{\mathcal{S}_{HH}}=\left[\mathcal{S}^{f_k,\theta_1,\mathrm{HH}} \quad  \mathcal{S}^{f_k,\theta_2,\mathrm{HH}} \quad \cdots \quad \mathcal{S}^{f_k,\theta_{K_\theta},\mathrm{HH}}\right]^T$ and $\mathbf{\mathcal{S}_{VV}}=\left[\mathcal{S}^{f_k,\theta_1,\mathrm{VV}} \quad  \mathcal{S}^{f_k,\theta_2,\mathrm{VV}} \quad \cdots \quad \mathcal{S}^{f_k,\theta_{K_\theta},\mathrm{VV}}\right]^T$. The observation space dimension is  $4 \cdot K_\theta$.

\subsubsection{Likelihood model} $p(\mathbf{y} |\mathbf{x})$ describes the probabilistic relation between the system state $\mathbf{x}$ and the observation $\mathbf{y}$ (at frequency $f_k$). In other words, it provides the probability distribution of the observation  $\mathbf{y}$ given a known system state $\mathbf{x}$. It is a key element of the knowledge that needs to be taken into account. Our inference goal is going to inverse this statistical relation. The likelihood model can be expressed as  a multidimensional Gaussian of mean  $\mathcal{F}_\mathrm{Maxwell}(\mathbf{x})$ and covariance matrix $\mathbf{R_m}$: 
\begin{equation}
\mathbf{y} | \mathbf{x} \sim \mathcal{N}(\mathcal{F}_\mathrm{Maxwell}(\mathbf{x}),\mathbf{R_m}) \label{Lhmod}
\end{equation}

where $\mathcal{F}_\mathrm{Maxwell}$ is the direct model, from the state space to the observation space, that relies on the aforementioned Maxwell solver. Taking into account measurement uncertainties, the likelihood model  results from the following considerations.
\begin{itemize}
\item[-]  The Maxwell solver, based on a direct method, is exact, i.e. extremely precise. $\mathcal{F}_\mathrm{Maxwell}$ is assumed to compute the "perfect observations", meaning without measurement noise, bias, etc. Implicitly, it is assumed that the shape object is perfectly known and that, conditionally to radioelectric properties, uncertainty only comes from measurement.
\item[-] From previous measurement uncertainty analysisis (see metrology guideline \cite{IEEEstd}), it has been  shown that the measurement uncertainty can be reasonably modeled by an additive Gaussian noise ($\mathbf{y}=\mathcal{F}_\mathrm{Maxwell}(\mathbf{x})+\mathbf{v_m}$, $\mathbf{v_m} \sim \mathcal{N}(\mathbf{0},\mathbf{R_m})$) with the quantified  covariance matrix $\mathbf{R_m}$.
\end{itemize}

\paragraph*{}
Consequently, the likelihood model can be expressed as (with $\nu=4 \cdot K_\theta$):
\begin{equation}
p(\mathbf{y} |\mathbf{x})=\frac{1}{(2\pi)^{\frac{\nu}{2}} \sqrt{\det \mathbf{R}}}e^{-\frac{1}{2} (\mathbf{y}-\mathcal{F}_\mathrm{Maxwell}(\mathbf{x}))^T \mathbf{R}^{-1} (\mathbf{y}-\mathcal{F}_\mathrm{Maxwell}(\mathbf{x}))}
\end{equation}
At first sight, just considering a single frequency $f_k$, numerous evaluations of $p(\mathbf{y} |\mathbf{x})$, i.e. of the Maxwell solver $\mathcal{F}_\mathrm{Maxwell}(\mathbf{x})$, are required in order to solve the inverse problem; they can be far too time-consuming, even with high performance computing. To avoid  heavy $\mathcal{F}_{\mathrm{Maxwell}}$ computations, a statistical learning approach has been achieved.  Its basic principle is to build a surrogate model, i.e. an approximation of  $\mathcal{F}_{\mathrm{Maxwell}}$ that is acceptable in the limited domain of interest. In a way, it is related to weak scattering linearization approximation methods of \cite{colton2003linear} in inverse scattering, and among them, the widely used Born approximation  \cite{hopcraft1992introduction,colton2003linear}. Here, the statistical linearization is not performed from truncation of physical interactions, but from full Maxwell solution computations that take into account multiple interactions, creeping waves, etc. The system, i.e. the high dimension state space of $\mathbf{x}$ and the associated system response $\mathcal{F}_\mathrm{Maxwell}(\mathbf{x})$, is explored by random sampling, according  to a prior knowledge about the expected radioelectric properties (prior distribution $p(\mathbf{x})$).  The computations are massively distributed on HPC machines, each computation involving the parallelized Maxwell solver. The computation number depends mainly on the state space dimension. The Monte Carlo simulation process leads to the following training set: 
\begin{equation} 
\mathcal{B}={\{(\mathbf{x}^{(1)},\mathbf{y}^{(1)}),(\mathbf{x}^{(2)},\mathbf{y}^{(2)}),\cdots, (\mathbf{x}^{(N_S)},\mathbf{y}^{(N_S)})\}}
\end{equation}
where $\mathbf{x}^{(k)} \sim p(\mathbf{x})$ ($\sim$ for realization of) and $\mathbf{y}^{(k)}=\mathcal{F}_\mathrm{Maxwell}(\mathbf{x}^{(k)})$ (for $k=1 \cdots N_S$), $N_S$ being the number of samples. Multidimensional linear regression provides a straightforward and efficient way to build a linear model $\mathbf{y}=f(\mathbf{x})+\mathbf{v_l}$ ($\mathbf{v_l}$ is an linearization error term) with:
\begin{equation}
f(\mathbf{x})=\mathbf{A} \cdot \mathbf{x} + \mathbf{y}^0  \mbox{ or }  f(\mathbf{x})=\mathbf{A}^\star \cdot [1 \quad \mathbf{x}] \mbox{ , }  \mathbf{A}^\star=[\mathbf{y}^0 \quad \mathbf{A}]
\end{equation}
$\mathbf{A}^\star$ is the least square (LS) estimates of the matrix of parameters that minimizes the errors to linearity ($\delta_l$), is given by the solution to the normal equations: 

\begin{equation}
\mathbf{A}^\star=(\mathcal{X}_\mathcal{B}^T\cdot\mathcal{X}_\mathcal{B})^{-1}\mathcal{X}_\mathcal{B}^T\mathcal{Y}_\mathcal{B} 
\mbox{ with } \mathcal{X}_\mathcal{B}=\left[\begin{array}{cc} 1 & \mathbf{x}^{(1)} \\ 1 &  \mathbf{x}^{(2)} \\ \cdots & \cdots \\ 1 &  \mathbf{x}^{(N_S)} \end{array} \right]
\mbox{, } \mathcal{Y}_\mathcal{B}=\left[\begin{array}{c} \mathbf{y}^{(1)} \\ \mathbf{y}^{(2)} \\  \cdots \\ \mathbf{y}^{(N_S)} \end{array} \right]
\end{equation}
where  $\mathcal{X}_\mathcal{B}$ is the ($4 N\times N_S$)  input  matrix and  $\mathcal{Y}_\mathcal{B}$ the ( $4 K_\theta \times N_S$)  response matrix,  from the training set $\mathcal{B}$. For numerical stability, a QR decomposition of $\mathcal{X}_\mathcal{B}$ is introduced. By residual analysis, it is then possible to assess the linear model fitness, i.e. to determine the discrepancy between the data and the model in the domain of interest. In principle, the covariance matrix ($\mathbf{R}_l$) evaluation of the linearization error $\mathbf{v_l}$ may require a supplementary  data set or cross-validation methods. Remark that additional statistical analysis can be achieved to extract reduced models, removing useless explanatory variables, i. e. permittivity or permeability components of zone subsets. That depends on the wave interaction, especially on the frequency band.

\begin{figure}[ht!]
 \centerline{ \includegraphics[width=0.4\columnwidth,clip=true,viewport=0cm 0cm 20cm 17.0cm,draft=false]{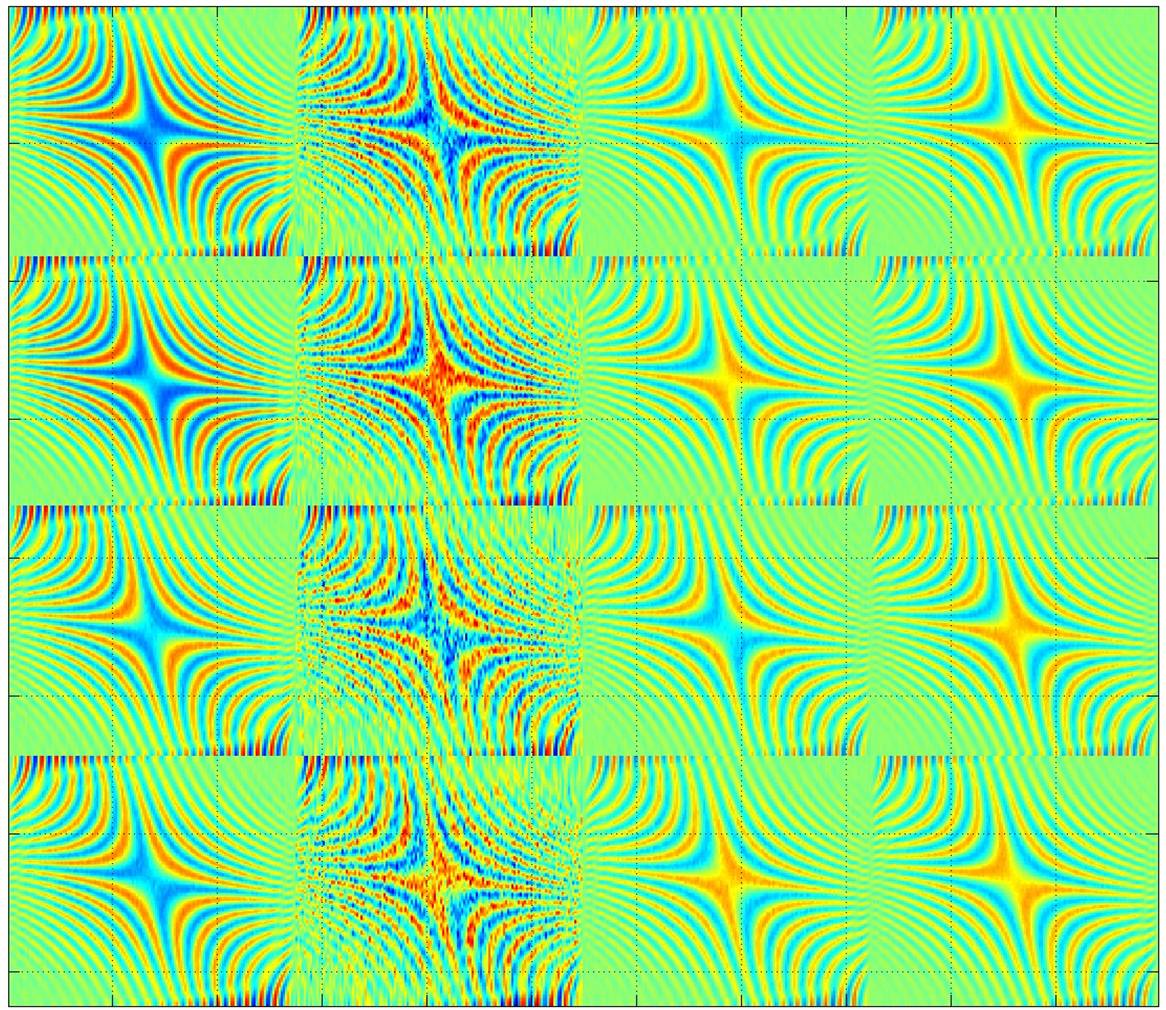}} 
\caption{Matrix $\mathbf{A}_k$ illustration} \label{A_OgBANG} 
\end{figure}

\paragraph*{}
Back to the likelihood model (\ref{Lhmod}),  it leads to an overall error term $\mathbf{v}=\mathbf{v_l}+\mathbf{v_m}$ of covariance matrix $\mathbf{R}$ \footnote{In our context, the linearization error turns out to be much lesser than the RCS measurement uncertainties: $\mathbf{R}+\mathbf{R}_l \simeq \mathbf{R}$.} and to the following linear Gaussian (LG) likelihood model (reintroducing the subscript $k$ for frequency $f_k$):

\begin{equation}
\mathbf{y}_k | \mathbf{x}_k \sim \mathcal{N}(\mathbf{A}_k \cdot \mathbf{x}_k + \mathbf{y}^0_k,\mathbf{R}_k) 
\mbox{ or }
\mathbf{y}_k=\left[\mathbf{A}_k \cdot \mathbf{x}_k + \mathbf{y}_k^0 \right]+\mathbf{\mathbf{v}_k}
\label{Lhmod1}
\end{equation}
with $\mathbf{A}_k$ and $\mathbf{y}^0_k$ learned from the training set $\mathcal{B}_k$. It is illustrated in figure \ref{A_OgBANG} for the ogival shape of figure \ref{Ogive_Areas} ($N=137$, $f=1.5$ GHz, $\theta=0^\circ:1^\circ:180^\circ$ - exploration: $1000$ HPC $\mathcal{F}_{\mathrm{Maxwell}}$ simulations). Inside each bloc, the pattern can be explained by the coherent contribution of each elementary zone. 

\subsubsection{Bayesian approach}
If such an inversion at a single frequency $f_k$ could be solved by classical regularization methods \cite{demoment1989image},  Bayesian estimation offers a convenient and powerful framework. Let us probabilize the unknown state vector $\mathbf{x}_k$ and consider a prior probability distribution $p(\mathbf{x}_k)$. It is possible to model the priori knowledge with a Gaussian distribution: $\mathbf{x}_k \sim \mathcal{N} \left( \mathbf{m}_k , \mathbf{P}_k \right)$. 

\begin{description}
\item[The mean $\mathbf{m}_k$] (dimension $N$) defines the reference radioelectric properties for the $N_a$ areas that divide the object (cf. figure \ref{Ogive_Areas}). 

\begin{equation}
\mathbf{m}_k=\left[\mathbf{m}^{\epsilon^{\prime}}_k  \quad \mathbf{m}^{\epsilon^{\prime\prime}}_k  \quad \mathbf{m}^{\mu^{\prime}}_k  \quad \mathbf{m}^{\mu^{\prime\prime}}_k\right]^T
\label{def_mean}
\end{equation}

where $\mathbf{m}^{\epsilon^{\prime}}_k=[\underbrace{\epsilon^{\prime}_k(1) \cdots \epsilon^{\prime}_k(1)}_{\mbox{area }1}  \quad  \underbrace{\epsilon^{\prime}_k(2) \cdots \epsilon^{\prime}_k(2)}_{\mbox{area } 2} 
\quad \cdots \quad \underbrace{\epsilon^{\prime}_k(N_a) \cdots \epsilon^{\prime}_k(N_a)}_{\mbox{area } N_a}]^T$,  $\epsilon^{\prime}_k(i)$  being the reference real permittivity of area $i$ ($i=1\cdots N_a$). Similar construction for $\mathbf{m}^{\epsilon^{\prime\prime}}_k$, $\mathbf{m}^{\mu^{\prime}}_k$ and $\mathbf{m}^{\mu^{\prime\prime}}_k$.

\item[The covariance $\mathbf{P}_k$] (dimension $N \times N$) quantifies the prior uncertainty around $\mathbf{m}_k$. $\mathbf{P}_k$ is block-diagonal: $
\mathbf{P}_k=\mathtt{diag}(\mathbf{P}^{\epsilon^{\prime}}_k,\mathbf{P}^{\epsilon^{\prime\prime}}_k,\mathbf{P}^{\mu^{\prime}}_k,\mathbf{P}^{\mu^{\prime\prime}}_k)
$. It means that the properties $(\epsilon^{\prime}$, $\epsilon^{\prime \prime}$, $\mu^{\prime}$, $\mu^{\prime \prime})$ are assumed to be uncorrelated. Each property block is block-structured itself. For instance, 
$\mathbf{P}^{\epsilon^{\prime}}_k=\mathtt{diag}(\mathbf{P}^{\epsilon^{\prime}}_k(1), (\mathbf{P}^{\epsilon^{\prime}}_k(2), \cdots,(\mathbf{P}^{\epsilon^{\prime}}_k(N_a))$, expressing the assumed property independence between areas. Focusing on one block $\mathbf{P}^{\epsilon^{\prime}}_k(i)$, a  squared exponential covariance expresses the spatial homogeneity (of the given property) between components, i.e. elementary zones of the object that belong to the same $i^{th}$ material area :

\begin{equation}
\mathbf{P}^{\epsilon^{\prime}}_k(i)=\left[\sigma^{\epsilon^{\prime}}_k(i)\right]^2 \times \left[\begin{array}{ccccc}
1 & \rho_S & \rho_S^2 & \cdots & \rho_S^{n-1}  \\
\rho_S & 1  & \rho_S &  &  \vdots \\
\rho_S^2 & \rho_S & \ddots & \ddots & \vdots \\
\vdots &  &  \ddots & \ddots &  \rho_S  \\
\rho_S^{n-1} & \cdots & \cdots & \rho_S & 1
                        \end{array} \right]
\label{def_cov} 
\end{equation}
with $\left[\sigma^{\epsilon^{\prime}}_k(i)\right]^2$ is the spatial variance of $i^{th}$  area and $\rho_S \in [0,1]$ the normalized spatial correlation parameter (e.g. $\rho_S = 0.95$). With this Markovian property,  commonly used in Gaussian field modeling, correlation decreases geometrically with the distance between components. Similar construction for $\mathbf{P}^{\epsilon^{\prime\prime}}_k$, $\mathbf{P}^{\mu^{\prime}}_k$ and $\mathbf{P}^{\mu^{\prime\prime}}_k$.

\end{description}
With linear Gaussian structure, i.e. Gaussian prior and  linear Gaussian likelihood, Bayesian inversion can be performed straightforwardly, with closed-form solutions.
In our problem, it is a piece of the more complex global problem that encompasses the frequency variation.

\subsection{The global problem statement} 

Radioelectric properties are known to vary in function of the wave frequency \cite{knott2004radar}.  They can be quite different from  the lower band frequency $f_1$ to the higher band one $f_K$. The basic idea is to maintain the former statistical modeling at each frequency $f_k$ while introducing additional a priori information about the dynamic in frequency, i.e. how quickly can a property move with frequency, how correlated are a property at two different frequencies, etc. This regularity information can be quite different from one EM property ($\epsilon^{\prime}$, $\epsilon^{\prime \prime}$, $\mu^{\prime}$, $\mu^{\prime \prime}$) to another, as well as from one material to another, 

\subsection{Generalized Auto-Regressive random process}
The statistical modeling extension consists in modeling the whole sequence $(\mathbf{x}_k, k \in \{1, \ldots , K_f \})$ by a generalized autoregressive (AR) random process:

\begin{eqnarray}
  \mathbf{x}_1   & \sim \ \mathcal{N} \left( \mathbf{m}_1, \mathbf{P}_1\right)  \nonumber \\
  \mathbf{x}_{k+1}  & =  \mathbf{m}_{k+1}+\mathbf{D}_{\rho} \cdot \mathbf{H}_{k+1} \cdot \mathbf{H}_k^{-1} \cdot \left( \mathbf{x}_k - \mathbf{m}_k \right) + \sqrt{\mathbf{I_d}-\mathbf{D}_{\rho}^2} \cdot \mathbf{H}_{k+1} \cdot  \mathbf{V}_k
  \label{mod-AP}
\end{eqnarray}

where $\mathbf{H}_k$ is the square root of the covariance matrix $\mathbf{P}_k$  \footnote{ unique symmetric definite positive matrix such as: $\mathbf{H}_k \cdot \mathbf{H}_k^T = \mathbf{P}_k$.}. $(\mathbf{V}_k,k \in \{1, \ldots , K \})$ are i.i.d. (independent, identically distributed) $\mathcal{N} (0,\mathbf{I_d})$ and $\mathbf{D}_{\rho}$ is a positive diagonal matrix commuting with $\mathbf{H}_k$. The dynamic model expresses the linear Gaussian correlation structure. It can be checked that the marginal distribution of $\mathbf{x}_k$ is still $\mathcal{N} \left( \mathbf{m}_k , \mathbf{P}_k \right)$. More generally, it can be shown that the distribution of concatenated vector $\mathbf{x}=(\mathbf{x}_1,\ldots,\mathbf{x}_{K_f})$ is Gaussian  with mean $\mathbf{m}=(\mathbf{m}_1,\ldots,\mathbf{m}_{K_f}) $ and covariance matrix:
\begin{equation}
\mathbf{P}  =  \mathbf{\mathcal{H}} \cdot
\left[ \begin{array}{ccccc}
\mathbf{I_d} & \mathbf{D}_{\rho} & \mathbf{D}_{\rho}^2 & \cdots & \mathbf{D}_{\rho}^{K_f-1}  \\
\mathbf{D}_{\rho} & \mathbf{I_d}  & \mathbf{D}_{\rho} &  &  \vdots \\
\mathbf{D}_{\rho}^2 & \mathbf{D}_{\rho} & \ddots & \ddots & \vdots \\
\vdots &  &  \ddots & \ddots &  \mathbf{D}_{\rho}  \\
\mathbf{D}_{\rho}^{K_f-1} & \cdots & \cdots & \mathbf{D}_{\rho} & \mathbf{I_d}
                        \end{array} \right]
\cdot \mathbf{\mathcal{H}}^T \label{ModAR}
\end{equation}

where $\mathbf{\mathcal{H}}$ is the block diagonal matrix $\mathbf{\mathcal{H}}=\mathtt{diag} (\mathbf{H}_1, \ldots, \mathbf{H}_{K_f}) $. Basically,  every joint distribution $(\mathbf{x}_i , \mathbf{x}_j)$ is expressed .

\paragraph*{}
The matrix $\mathbf{D}_{\rho}$ takes into account the frequential  correlations of the EM properties $\mathbf{x}_1 \cdots \mathbf{x}_{K_f}$; it refers to a hyper-parameter $\rho$. According to about frequency correlation prior knowledge, the following alternatives can be considered:
\begin{enumerate}
 \item The frequency correlation doesn't depend on the material and  the EM property ($\epsilon^{\prime}$, $\epsilon^{\prime \prime}$, $\mu^{\prime}$ or $\mu^{\prime \prime}$): $\rho$ is scalar ($\in [0,1]$) and $\mathbf{D}_{\rho} = \rho.\mathbf{I_d}$.
 \item  It depends on the material: $\rho$ is $N_a$-dimensional ($\in [0,1]^{N_a}$), and $\mathbf{D}_{\rho}$ is the block-diagonal matrix made up of $N_a$ terms $\rho_i . \mathbf{I_d}$.
 \item  It depends on both:  $\rho$ is $4.N_a$-dimensional and $\mathbf{D}_{\rho}$ is the block-diagonal matrix made up of $4 .N_a$ terms $\rho_i . \mathbf{I_d}$.
\end{enumerate}

\subsection{A conditionally hidden dynamic Markov process}
The generalized AR random processes include the linear Gaussian models at the various frequencies $f_k$ ($k=1 \cdots K_f$). It provides a spatial and frequential correlation structure. Assuming that the material areas are known to be quite homogeneous, the spatial correlation parameter can be fixed (typically $\rho_S=0.95$). Quite the reverse,  frequency correlations can not be really known; they are to be determined by the inversion process. Back to Bayesian statistics, it is chosen to probabilize the unknown hyper-parameter $\rho$. Finally, the combination of the AR dynamic model (\ref{ModAR}) and the likelihood model (\ref{Lhmod1}) end in the following state-space model, observed at "times" $f_k$ ($k=1,\cdots,K$):
\begin{equation}
 	\mathbf{x}_{k+1}=\mathbf{M}_k^{\rho}\cdot\mathbf{x}_k+\mathbf{w}_k \qquad
 	\mathbf{y}_k=\left[\mathbf{A}_k \cdot \mathbf{x}_k + \mathbf{y}_k^0 \right]+\mathbf{v}_k
\end{equation}
assuming the initial state $\mathbf{x}_1 \sim \mathcal{N}(\mathbf{m_1},\mathbf{P_1})$. $\mathbf{M}_{k}^{\rho}$ is a transition matrix and $\mathbf{w}_k$ a Gaussian model  noise ($\mathbb{E}(\mathbf{w}_k)\neq0$). Both directly arise from (\ref{ModAR}); they are not detailed here for clearness. 

\begin{figure}[ht!]
 \centerline{ \includegraphics[width=0.5\columnwidth,clip=true,viewport=5cm 5cm 22cm 16cm,draft=false]{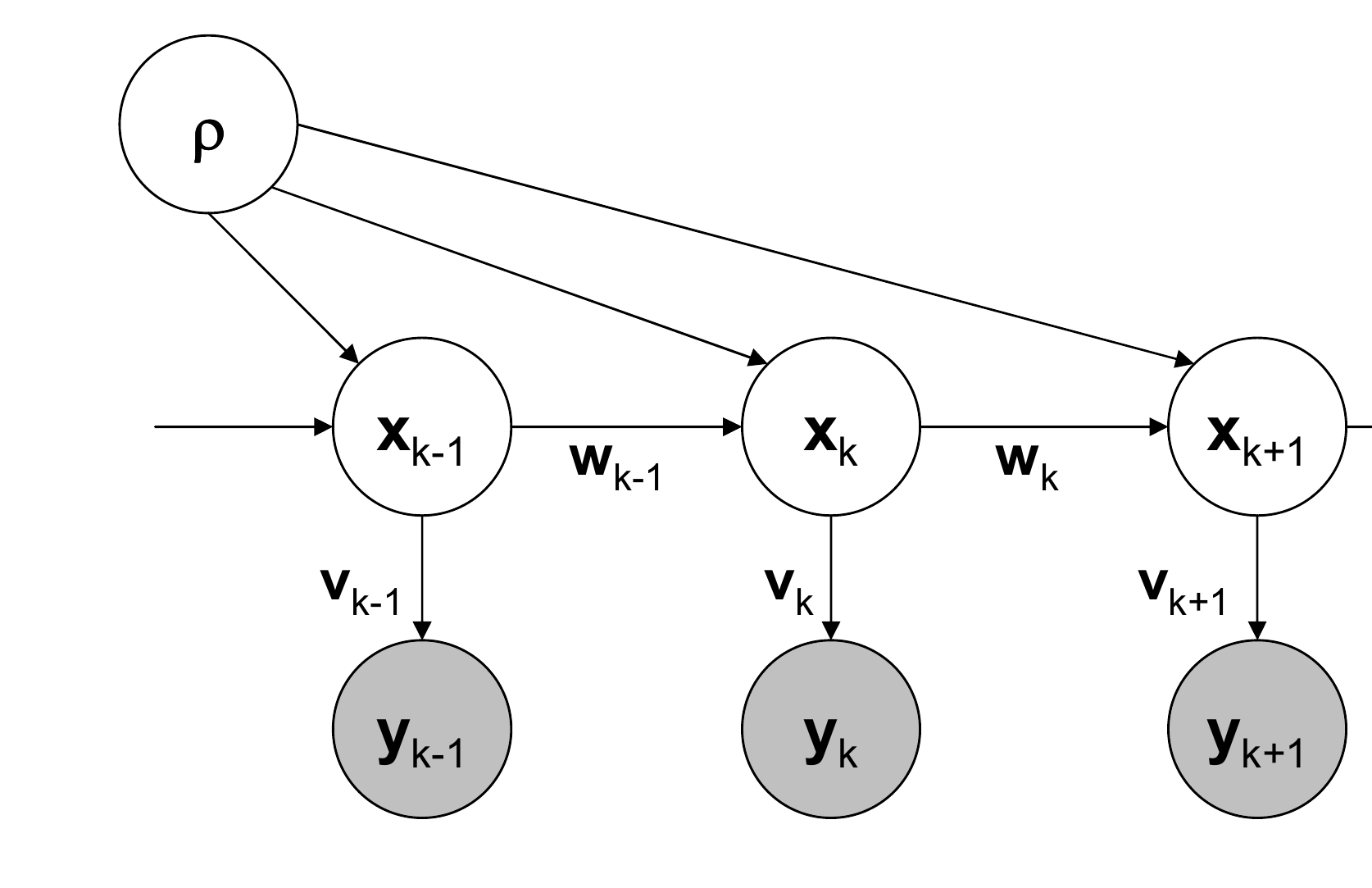}} 
\caption{A graphical representation} \label{DAG} 
\end{figure}

Again, let emphasize that the dynamic model involves that each marginal complies with  $\mathbf{x}_{k} \sim \mathcal{N}(\mathbf{m_k},\mathbf{P_k})$. On the other hand, it is important to remark that, conditionally to the frequential correlation parameter $\rho$, the model is a classic linear Gaussian hidden dynamic Markov process. A graphical representation of the entire model is given in figure \ref{DAG}. Given a value of $\rho$, the lower part describes a linear gaussian system. The idea is to make the most of this specific structure.



\section{Advanced Sequential Monte Carlo inversion} \label{sec-traitement}

\subsection{The Rao-Blackwellized Approach}

As already mentioned, the unknown hyper-parameter $\rho$ is probabilized, so it is given a prior distribution $p(\rho)$, assumed calculable (up to a normalizing constant) and easy to sample. The posterior distribution $p(\mathbf{x},\rho| \mathbf{y})$ can be decomposed as:
\begin{equation}
p(\mathbf{x},\rho| \mathbf{y}) =  p(\mathbf{x}| \rho,\mathbf{y}) \cdot p(\rho|\mathbf{y})
\end{equation}

Since  the system is linear Gaussian conditionally to $\rho$, the conditional distributions $p(\mathbf{x}_k| \rho,\mathbf{y})$ can be straightforwardly computed by classic Kalman filtering. This forward algorithm can be completed by backward smoothing, in this off-line context;  the overall is often called "Kalman smoother". On the other hand, the term $p(\rho|\mathbf{y})$ can been decomposed as:
\begin{equation}
\begin{array}{rcl}
p(\rho|\mathbf{y})  &  \propto &  p(\rho) \cdot p(\mathbf{y}| \rho)  \\
                    &  \propto & \displaystyle{ p(\rho) \cdot \prod_{k=1}^{K_f} \underbrace{ p \left( \mathbf{y}_k | \rho, \mathbf{y}_1, \ldots , \mathbf{y}_{k-1} \right)  }_{:=J_k(\rho)}  }  .            
\end{array}
\end{equation}

Again, for any hyper-parameter $\rho$, the quantities $J_k(\rho)$ can be evaluated from the likelihood terms provided by the Kalman filter.
Eventually, it is possible to exploit this conditional system structure, with Kalman smoothers that can be applied and integrated in the following interacting particle approach. In a first step, a stochastic algorithm (described in section \ref{sec-descr-SMC}) gives an approximation of $p(\rho|\mathbf{y})$. It estimates the frequential correlations (i.e. regularity) of the EM properties $\epsilon^{\prime}(f), \epsilon^{\prime \prime}(f), \mu^{\prime}(f), \mu^{\prime \prime}(f)$. In a second step, the first moments of $\mathbf{x}_k$ can be evaluated (for each frequency $f_k$) by the theoretical conditioning relations:
\begin{equation}
\begin{array}{l}
\mathbb{E} (\mathbf{x}_k | \mathbf{y})    =     \mathbb{E} \left[ \mathbb{E} (\mathbf{x}_k | \rho,\mathbf{y}) | \mathbf{y} \right]   \label{esp-RB}
\end{array} 
\end{equation}
\begin{equation}
\begin{array}{l}
  \mathbb{V}\mbox{ar} (\mathbf{x}_k | \mathbf{y}) =  \mathbb{E} \left[ \mathbb{V}\mbox{ar} (\mathbf{x}_k | \rho,\mathbf{y})  | \mathbf{y} \right]
+ \mathbb{V}\mbox{ar} \left[ \mathbb{E} (\mathbf{x}_k | \rho,\mathbf{y})  | \mathbf{y} \right]     \label{cov-RB}
\end{array} 
\end{equation}
Note that Kalman recursions are used both in the first step for calculating the likelihood of the hyper-parameter $\rho$ (up to a normalizing constant) and in the second step for determining the quantities $\mathbb{E} (\mathbf{x}_k | \rho,\mathbf{y})$ and $\mathbb{V}\mbox{ar} (\mathbf{x}_k | \rho,\mathbf{y})$. This idea of mixing analytic integration (here Kalman evaluation of $p(\mathbf{x}|\rho, \mathbf{y})$) with stochastic sampling (here to approximate $ p(\rho|\mathbf{y})$) is a variance reduction approach, known as Rao-Blackwellisation \cite{Liu}.\\

Let us  denote by $\eta (d \rho)$ the probability measure associated with the marginal distribution $p(\rho | \mathbf{y})$, for a fixed observation vector $\mathbf{y}$. Similarly to \cite{Liu}, we choose to implement for the first step an efficient interacting particle approach, called Sequential Monte Carlo (SMC), in order to estimate $\eta$. We now give a brief but general description of these methods.

\subsection{The SMC algorithm}  \label{sec-descr-SMC}
\begin{figure}[h]
\centering{\includegraphics[width=0.6\columnwidth]{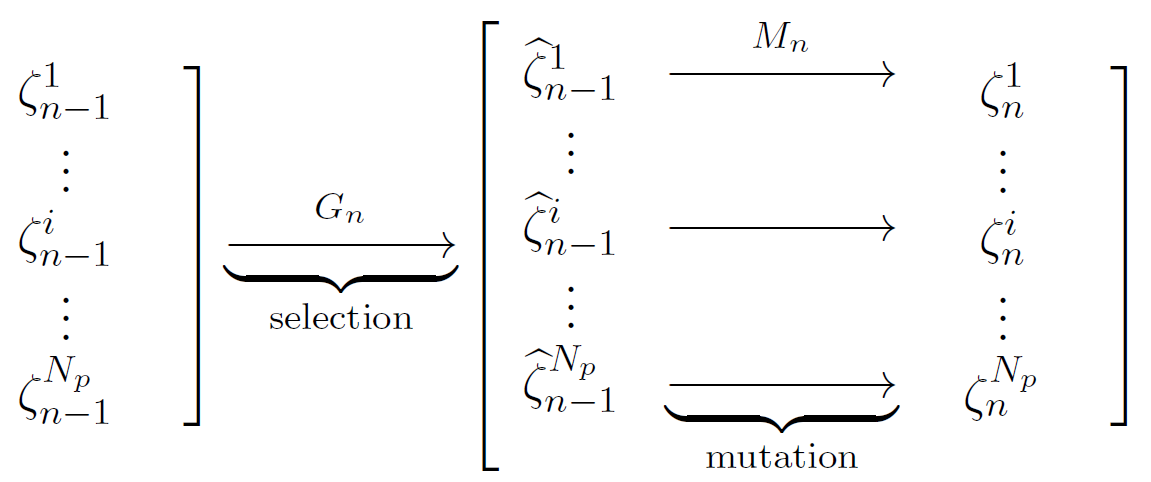}
\caption{\label{schema-SMC} The SMC 2-step evolution}}
\end{figure}

Sequential Monte Carlo is a stochastic algorithm to sample from complex high-dimensional probability distributions. The principle (see, e.g.,~\cite{DMDJ}) is to approximate a sequence of target probability distributions $(\eta_n)$ by a large cloud of random samples termed particles $(\zeta_{n}^k)_{1 \leq k \leq N_p} \in E^{N_p}$, $E$ being called the state space. Between ``times'' $n-1$ and $n$, the particles evolve in the state space $E$ according to two steps (see figure \ref{schema-SMC}):
\begin{enumerate}
 \item \textbf{A selection step}: every particle $\zeta_{n-1}^{i}$ is given a weight $\omega_i$ defined by a selection function $G_n: E \rightarrow (0,+\infty)$ (i.e. $\omega_i = G_n(\zeta_{n-1}^{i})$). By resampling (stochastic or deterministic), low-weighted particles vanish and are replaced by replicas of high-weighted ones.
 \item \textbf{A mutation step}: each selected particle $\widehat{\zeta}_{n-1}^{i}$ moves, independently from the others, according to a Markov kernel $M_n: E \rightarrow E$.
\end{enumerate}

\noindent Evolving this way, the cloud of particles, and more precisely the occupation distribution $\eta_n^{N_p} := \frac{1}{N_p} \sum_{k=1}^{N_p} \delta_{\zeta_k^n}$ (sum of Dirac distributions), approximates for each $n$ the theoretical distribution $\eta_n$ defined recursively by the Feynman-Kac formulae. It is associated with the potentials $G_n$ and kernels $M_n$ (see~\cite{DM-FK} for further details). More precisely, this sequence $\eta_n$ is defined by an initial probability measure $\eta_0$ and the recursion:
\begin{equation}
\quad \quad \eta_{n} = \Psi_{G_n} (\eta_{n-1})  .  M_n  \label{formule-FK}
\end{equation}
where $\Psi_{G_n} (\eta_{n-1})$ is the probability measure defined by $\Psi_{G_n} (\eta_{n-1}) (dx) \propto G_n(x).\eta_{n-1} (dx)$
and, for any probability measure $\mu$, $\mu.M_n$ is the measure so that $\mu.M_n (A) = \int_{E} M_n(x,A) \mu(dx)$.

The SMC approach is often used for solving sequential problems, such as filtering (e.g., \cite{Cappe,Doucet-F-G,DM-filt}). In other problems, like ours, this algorithm also turns out to be efficient to sample from a single target measure $\eta$. In this context, the central idea is to find a judicious interpolating sequence of probability measures $(\eta_n)_{0\leq k\leq n_f}$ with increasing sampling complexity, starting from some initial distribution $\eta_0$, up to the final target one $\eta_{n_f}=\eta$. Consecutive measures $\eta_n$ and $\eta_{n+1}$ are to be sufficiently similar to allow for efficient importance sampling and/or acceptance-rejection sampling. The sequential aspect of the approach is then an "artificial way" to solve gradually the sampling difficulty. More generally, a crucial point is that large population sizes allow to cover several modes simultaneously. This is an advantage compared to standard MCMC (Monte Carlo Markov Chain) methods that are more likely to be trapped in local modes. These sequential samplers have been used with success in several application domains, including rare events simulation (see \cite{Cerou-RE}), stochastic optimization and, more generally, Boltzmann-Gibbs measures sampling (\cite{DM-D-J}).

\subsection{Interpolating sequences of measures}  \label{sec-suite-mesures}
Back to our objective of sampling from $\eta (d \rho)$, let us denote by $E$ the state space of the variable $\rho$ (i.e. $E=[0,1]$, $[0,1]^{N_a}$ or $[0,1]^{4N_a}$). We have to define a sequence of distributions $(\eta_n)_{0\leq k\leq n_f}$ from the initial distribution $\eta_0(d \rho) = p(\rho) d \rho$ (easy to sample) to the target one $\eta_{n_f} (d \rho) = \eta(d \rho) = p(\rho | \mathbf{y}) d\rho$.

\subsubsection{The guiding principle}
 With this in mind, we first define an interesting class of Markov kernels on $E$: let $h$ be a positive, bounded function on $E$, and let $Q(x,dy)$ be a Markov kernel on $E$, assumed reversible w.r.t. the Lebesgue measure on $E$. The Metropolis-Hastings kernel $K_{h,Q}(x,dy)$ associated with $h$ and $Q$ is given by the following formula:
$$ 
    \begin{array}{llll}
         K_{h,Q}(x,dy)= Q(x,dy). \min \left( 1 , \frac{h(y)}{h(x)}   \right) & \forall y \neq x & \\
         & &\\
         K_{h,Q}(x,\{ x \}) = 1 - \underset{y \neq x}{\displaystyle{\int}}  Q(x,dy). \min \left( 1 , \frac{h(y)}{h(x)}   \right)         
    \end{array} 
$$
Using an acceptance/rejection method, this kernel is easy to sample as soon as one can sample $Q(x,dy)$ and calculate the ratios $h(y) / h(x)$. Here is a crucial property: if $\mu_h$ denotes the probability measure defined by $\mu_h(d \rho) \propto h(\rho) d\rho$, then it is well known (see, e.g.,~\cite{DM-Guionnet}) that $K_{h,Q}$ admits $\mu_h$ as an invariant measure:
$$
\mu_h.K_{h,Q} = \mu_h \quad \left( \Longleftrightarrow \underset{E}{\int} K_{h,Q}(\rho,A) \mu_h (d \rho)  =  \mu_h (A) \;,\; \forall A \subset E  \right)
$$
More generally, this property is satisfied for the iterated kernel $K_{h,Q}^m$, i.e. $\mu_h.K_{h,Q}^m = \mu_h$ (for any integer $m$).

Let $\eta_n$ be a sequence of probability measures defined with some positive, bounded functions $h_n$ so that: $\eta_n(d\rho) \propto h_n(\rho).d \rho$. Then, for any sequence of reversible Markov kernels $Q_n$ and any sequence of integers $m_n$, $\eta_n$ satisfies the Feynman-Kac formula (\ref{formule-FK}) with potentials $G_n:=h_n/h_{n-1}$ and Markov kernels $M_n := K_{h_n,Q_n}^{m_n}$ ($K_{h_n,Q_n}$ iterated $m_n$ times). Practically, the consequence  is that such a sequence $\eta_n$ can be approximated using a SMC algorithm as soon as one can calculate the functions $h_n$ up to a normalizing constant. Similarly to traditional MCMC or simulated annealing methods, this algorithm is all the more robust when the iteration numbers $m_n$ are large, since the kernels $K_{h_n,Q_n}$ are just defined and used to stabilize the system.\\

\subsubsection{Design of bridging measure sequences}
\noindent From these considerations, we propose three scheme variants of interpolating sequences of measures.
\begin{enumerate}
  \item The annealed scheme: the sequence $\eta_n$ is defined by the positive, bounded functions
$$
h_n(\rho) = p(\mathbf{y}| \rho)^{\alpha_n} \cdot p(\rho)
$$
where $(\alpha_n)_{1 \leq n \leq n_f}$ is a sequence of numbers increasing from $0$ to $1$ (arbitrarily chosen). In this situation, the potentials $G_n(\rho)$ used in the selection are equal to $p(\mathbf{y}| \rho)^{\alpha_n-\alpha_{n-1}}$. Thus, $\alpha_n$ is to be chosen to control the selectivity of these functions, which is important in practice. Annealing or tempering is frequently used in SMC (see \cite{jasra2007population,DMDJ}); it is related to simulated annealing (with inhomogeneous sequence of MCMC kernels). 

  \item The data tempered scheme: for all $n \in \{0,1,\ldots,K_f\}$, $\eta_n$ is the probability measure associated with: $h_n(\rho) = \displaystyle{p(\rho) \cdot \prod_{k=1}^{n} \underbrace{ p \left( \mathbf{y}_k | \rho, \mathbf{y}_1, \ldots , \mathbf{y}_{k-1} \right)  }_{=J_k(\rho)}}$.
In other words, at each generation $n$, the selection potential $G_n(\rho)$ that is applied to the particles is the term $p \left( \mathbf{y}_n | \rho, \mathbf{y}_1, \ldots , \mathbf{y}_{n-1} \right)$, i.e. the likelihood of the $n$-th observation vector given the previous ones. This allows the algorithm to work "online", since it treats sequentially the observations. According to \cite{jasra2007population}, it is efficient for problems that exhibit a natural order (e.g. hidden Markov models). Yet, when these potentials turn out to be too selective, the SMC algorithm turns out to perform poorly since the cloud of particles loses its diversity at each selection step. It is substituted for the next scheme that overcomes this drawback.

  \item The hybrid scheme: similarly to the previous one, this scheme incorporates the observations one after the other, but each likelihood function $J_k(\rho)$ is handled as a product:
  $$
  J_k(\rho) = \prod_{i=1}^{n_k} J_k(\rho)^{(\alpha^{(k)}_i-\alpha^{(k)}_{i-1})}
  $$
  where for all $k \in \{1,\ldots,K_f \}$, $(\alpha^{(k)}_{i})_{1 \leq i \leq n_k}$ is a sequence $0 \nearrow 1$. Then, if $n=(n_1 + \cdots + n_{r-1}) + s$, the function $h_n$ is given by:
  $$
  h_n(\rho) = p(\rho) \cdot \left( \prod_{k=1}^{r-1} J_k(\rho) \right) \cdot J_{r} (\rho)^{\alpha^{(r)}_s}
  $$
  Note that the selection potential $G_n = J_{r}^{(\alpha^{(r)}_s-\alpha^{(r)}_{s-1})}$ can be arbitrarily controlled.
\end{enumerate}

\noindent For each of these interpolating schemes, the functions $h_n$ are calculable up to a normalizing constant (Kalman equations), so that the Metropolis-Hastings kernels (possibly iterated) can be used to perform the mutation steps.

\subsection{The global estimation}

To sum up, the joint distribution $p(\mathbf{x},\rho| \mathbf{y})$ can decomposed and evaluated as follows:

\begin{equation*}
p(\mathbf{x},\rho| \mathbf{y}) =  \underbrace{p(\mathbf{x}| \rho,\mathbf{y})}_{\mbox{\tiny{KF (+ smoothing)}}} 
\cdot
\underbrace{
\overbrace{p(\rho|\mathbf{y})}^{
           \propto 
	\overbrace{p(\mathbf{y}| \rho)}^{\mbox{\tiny{ KF output}}}
	\cdot 
	 \overbrace{p(\rho)}^{\mbox{\tiny{prior}}}           
           }  
           }_{\mbox{\tiny{SMC}}}    
\end{equation*}
As previously mentioned, the SMC algorithm of section \ref{sec-descr-SMC} provides in the first stage an evaluation of the frequency correlations $p(\rho | \mathbf{y})$ (i.e. an approximation $\hat{\eta} = \eta_{n_f}^{N_p}$ of $\eta$). It is computed from the last generation of particles $(\rho^{(1)},\ldots,\rho^{(N_p)}) := (\zeta_{n_f}^1,\ldots,\zeta_{n_f}^{N_p})$. In the second stage, estimators of EM properties are straightforwardly computed from conditioning relations (\ref{esp-RB}) and  (\ref{cov-RB}) (see details in annex \ref{SecAnnexe}); it consists in approximations of the mean and covariance matrix of the system state $\mathbf{x}_k$. Focusing on a given frequency or on a fixed zone, the SMC method provides useful information:
\begin{itemize}
  \item[-] For any frequency $f_k$, it computes an approximation of the mean and covariance matrix of the system state $\mathbf{x}_k$. Roughly speaking, one can sample from the posterior distribution $p(\mathbf{x}_k|\mathbf{y})$ by picking a $\rho^{(i)}$ from the final cloud of particles and computing associated samples of $\mathbf{x}_k$ by a Kalman smoother conditionally to $\rho^{(i)}$ (see further illustration figure~\ref{freq-fix} page~\pageref{freq-fix}).
  \item[-] For any fixed zone, the method provides estimators of the mean and marginal variance for every frequency, so that the results can be presented as frequential profiles, with marginal uncertainties (using the diagonal values of  $\mathbf{\hat{\Sigma}}_k$) (see further illustration figure~\ref{ann-fix} page~\pageref{ann-fix}).
\end{itemize}


\section{Applications}    \label{sec-applications}
In this section, the inverse scattering approach is applied to EM scattering measurements of a metallic ogival shape object. The validation is achieved with simulated data in a wide frequency band from $f=200$ MHz to $8$ GHz. Section \ref{sec-scenario} describes the reference nondestructive testing scenario. Next, section \ref{sec-inversion} describes the inversion process and illustrates some results. A detailed performance analysis is developed in Section \ref{sec-performance}. Then, in Section \ref{sec-variantes}, we briefly analyze some variants of the approach.

\subsection{Nondestructive testing scenario}  \label{sec-scenario}

\paragraph{The metallic object} We consider the metallic axisymmetric object, previously shown in  figure \ref{Ogive_Areas}; its ogival shape, derived from the RCS benchmark \cite{woo1993programmer}, is perfectly known. The 2 m long object is coated by $N_a=5$ material areas, the isotropic radioelectric properties weakly varying within each area. For each material area, the true EM properties $\mathbf{x}_{\mbox{true}} (f)$ undergo the following model: 
$
\mathbf{x}_{\mbox{true}} (f) = \mathbf{x}_{\mbox{ref}} (f) + c \cdot \Lambda(f)
$.

\begin{figure}[h]
\centering{\includegraphics[width=0.7\columnwidth]{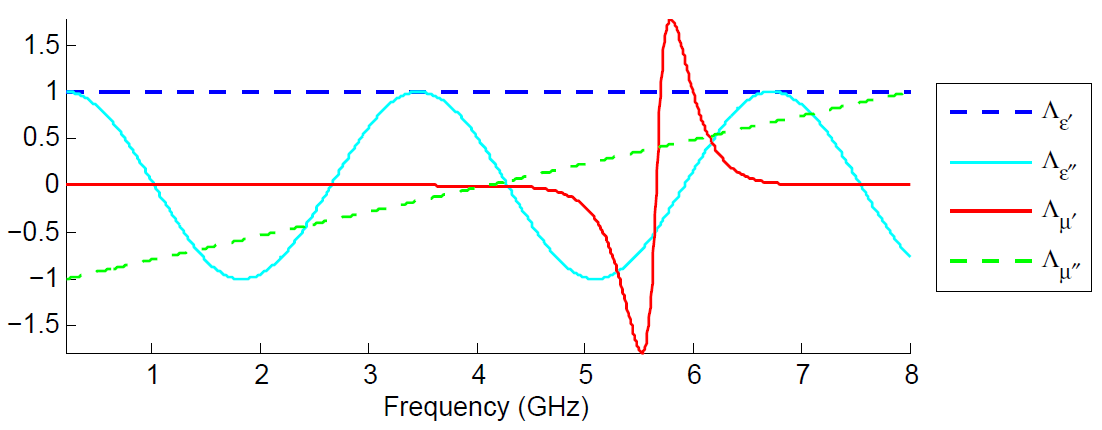}
\caption{\label{fonctions-lambda} The functions $\Lambda$ }}
\end{figure}

At each frequency $f$, the true (unknown) vector $\mathbf{x}_{\mbox{true}} (f)$  is $4N=76$-dimensional, where: 
\begin{itemize}
\item[-] $\mathbf{x}_{\mbox{ref}} (f)$ is a reference frequency profile, depending on the area and on the radioelectric component ($\epsilon^{\prime}, \epsilon^{\prime \prime}, \mu^{\prime}, \mu^{\prime \prime}$). Note that these $4N_a=20$ reference profiles are chosen regular and with typical orders of magnitude (i.e. non-negative and $\leq 20$).
\item[-] $\Lambda(f)$ is a perturbation function depending on the radioelectric component. Thus, the $4$ functions $\Lambda_{\epsilon^{\prime}}, \Lambda_{\epsilon^{\prime \prime}}, \Lambda_{\mu^{\prime}}, \Lambda_{\mu^{\prime \prime}}$ define the perturbation shapes . As shown in figure \ref{fonctions-lambda}, they are chosen more or less regular (in order to test the inversion capabilities).
\item[-] $c$ is a simple scaling factor, depending on the area. To examine the perturbation amplitude influence, increasing values of $c$ are chosen: $\{0.5,1,2,4,8\}$, related to the $5$ successiveareas.
\end{itemize}

\paragraph{(Simulated) scattering measurements}
According to the conventional RCS acquisition mode described in section \ref{SecPbEMmeas}, complex scattering coefficients are measured for both polarizations HH and VV, at $K_f=20$ regularly spaced frequencies  ($f_1 =0.2 \; \mbox{GHz},\cdots,f_{K_f}= 8 \;\mbox{GHz}$) and at $K_{\theta}=23$ regularly spaced incidence angles ($\theta_1=0^{\circ},\cdots, \theta_{K_{\theta}}=180^{\circ}$). 

\begin{figure}[h]
\centering{\includegraphics[width=0.7\columnwidth]{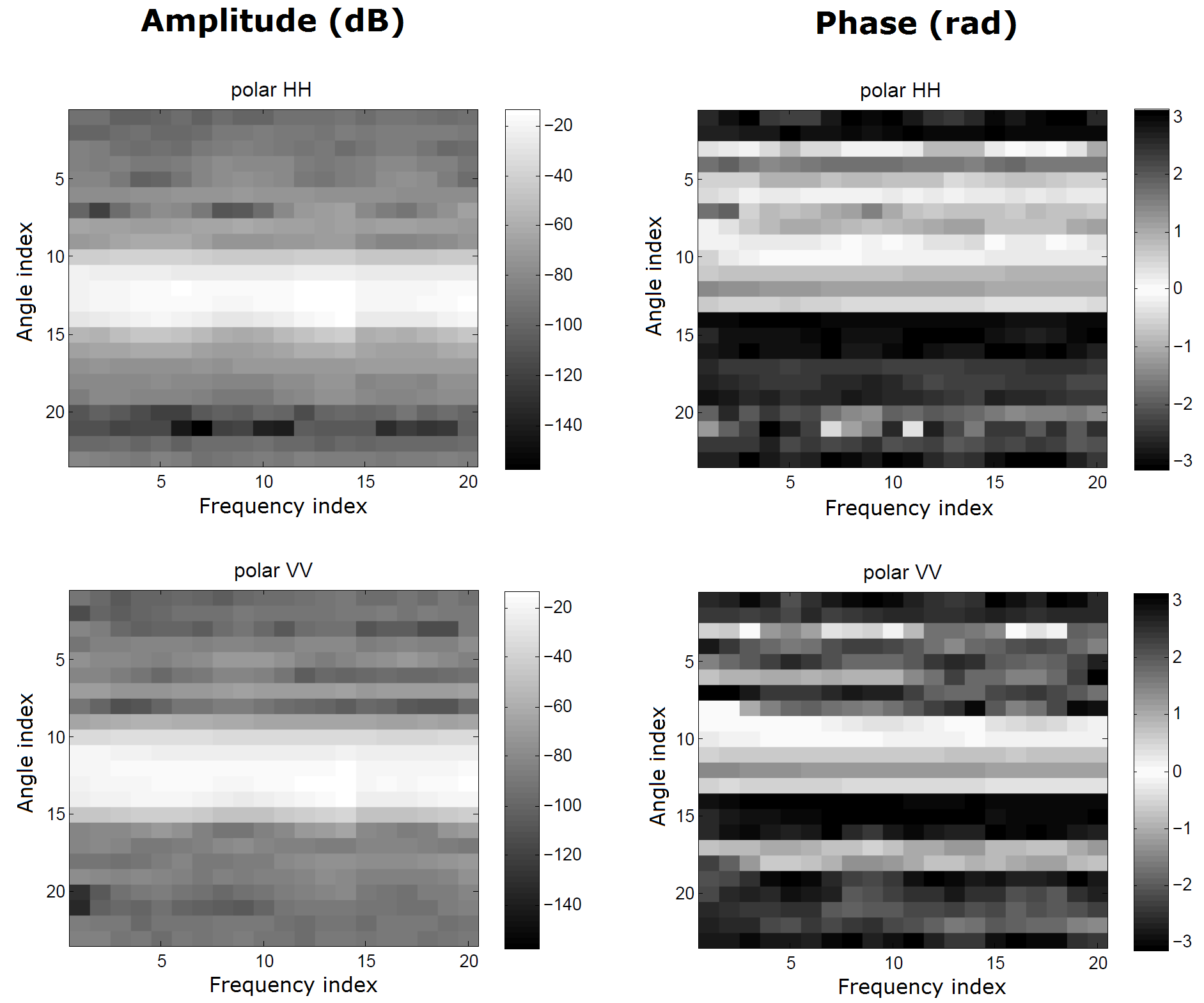}
\caption{\label{Fig_Obsdata} Observation hologram, amplitude and phase (polar HH and VV) }}
\end{figure}

The observation data $\mathbf{y}=(\mathbf{y}_1,\ldots,\mathbf{y}_{K_f})$ is simulated from the likelihood model (\ref{Lhmod}). That involves  to run the parallelized harmonic Maxwell solver ($\mathcal{F}_\mathrm{Maxwell}$) and to draw an additive white Gaussian noise of marginal standard deviation $\sigma_{\mathrm{n}} = 10^{-3}$ ($\sim$ $1\%$). Note that each of the $20$ observation vectors $\mathbf{y}_k$ is $4\times K_{\theta}=92$-dimensional. The data is represented in figure \ref{Fig_Obsdata}. Note on the amplitude representations the high specular reflections when the ogival object is turned perpendicularly to the wave propagation direction. 

\subsection{Inversion process}   \label{sec-inversion}
The goal is to estimate the radioelectric properties, the $\mathbf{x}_{\mbox{true}}$ term function of the frequency $f$, from the scattering measurements. In this section, we give a few implementation details regarding the application context.

\paragraph{State space} The state space dimension stems from the wave frequency number and from the discretization of the object in elementary mesh zones. In order to limit it, the cutting up of the object is here restricted to $N=19$ elementary zones. 

\begin{figure}[!h]
\centering{\includegraphics[width=0.55\columnwidth]{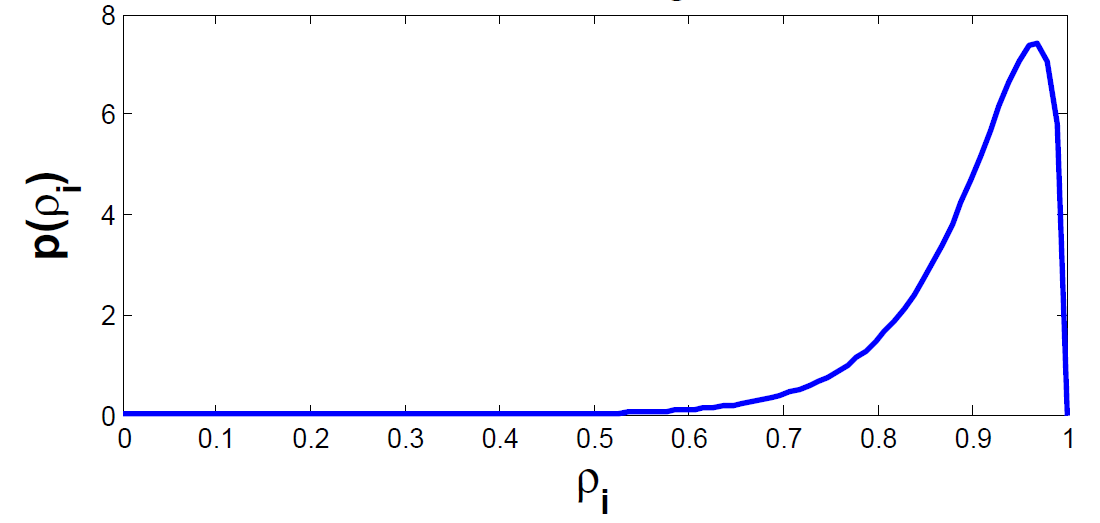}
\caption{\label{prior-rho} Marginal prior distribution $p(\rho_i)$}}
\end{figure}

\paragraph{Prior information} The prior information (see section \ref{SecMod}) needs to be detailed in this context. Concerning the prior spatial information $p(\mathbf{x}_k)$, its means $\mathbf{m}_k$ are given, for each $k$, by the former reference frequency profiles $\mathbf{x}_{\mbox{ref}} (f_k)$. Around them, the uncertainties are given by the block-structured covariance matrices $\mathbf{P}_k$ of (\ref{def_cov}) with: $\rho_{S} = 0.95$ and $\sigma_k (i)  =  1 + 0.15 \times \mathbf{m}_k(i)$ for any elementary zone $i$. In other words, we assume a minimum standard deviation of $1$ that increases proportionally to the reference amplitude value.
Regarding the prior frequential information, we assume that $\rho$ depends on both area and EM property ($\epsilon^{\prime}, \epsilon^{\prime \prime}, \mu^{\prime}, \mu^{\prime \prime}$), so that it is $20$-dimensional. As for its prior distribution $p(\rho)$, we set:
$$
p(\rho) = \prod_{i=1}^{20} p(\rho_i)
$$
where all the marginal prior distributions $p(\rho_i)$ are identical and presented on figure \ref{prior-rho}. Note that this distribution $p(\rho)$ can be sampled straightforwardly by sampling independently each component $\rho_i$ using, e.g., an acceptance/rejection method.

\paragraph{Likelihood model} The surrogate likelihood model (\ref{Lhmod1}) has been formerly learned:  $\mathbf{A}_k$ and $\mathbf{y}^0_k$ are known (see figure \ref{A_OgBANG}), as well as the marginal standard deviation $\sigma_{\mathrm{n}}$ which is in conformity with the measurement noise of the above observation simulation.

\paragraph{SMC tuning} The sequence of probability measures $\eta_n$ is standardly defined by the annealed scheme (see section \ref{sec-suite-mesures}). To ensure a stable behavior of the SMC algorithm (i.e. keep a good approximation $\eta_n^{N_p} \simeq \eta_n$ until the end), we chose the following efficient adaptive strategies (that make it possible to limit the number of particles to $N_p = 100$):
\begin{itemize}
\item[-] {\it selection step}: as mentioned, the increment ${\Delta \alpha}_n = \alpha_n - \alpha_{n-1}$ controls the selectivity degree. If ${\Delta \alpha}_n$ is too small, every particle is given approximately the same weight, and there is no selection among them. If ${\Delta \alpha}_n$ is too large, the majority of the particles are killed, the cloud loses all its diversity, and the SMC algorithm performs poorly. Therefore, instead of choosing beforehand ${\Delta \alpha}_n$, it is defined adaptively {\it so that} the selection step kills around $25 \%$ of the particle population. This is a way to ensure a reasonable selection.
\item[-] {\it mutation step}: the mutation step is crucial since it allows the particles to explore the state space $E$. We use Markov kernels $M_n$ defined as being the composition of several Metropolis-Hastings kernels $K_n^{(i)}$ whose proposition kernels $Q_n^{(i)} (x,dy)$ are uniform, centered in $x$, and associated with a window size $\sigma_{\mathrm{prop},n}^{(i)}$. To be sure that the particles move in a well-sized neighborhood, (i.e. large enough to explore $E$ and small enough to converge), the sequence $(\sigma_{\mathrm{prop},n}^{(i)})_i$ always starts with large values and decreases geometrically. Once more, we use an adaptive criteria to stop the process.
\end{itemize}

\begin{figure}[h]
\centering{\includegraphics[width=0.7\columnwidth]{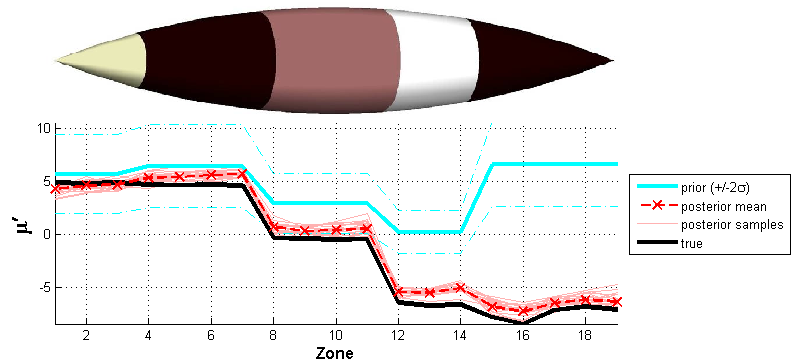}
\caption{\label{freq-fix} EM estimated properties at frequence $f=5.6 \; \mbox{GHz}$}}
\end{figure}

\paragraph{Results}
In the context of this reference study, the inversion process takes about $30$ minutes with a current standard processor. Note that the higher the dimension space is, the longer the inversion.  In figure \ref{freq-fix}, we show the estimations of $\mu^{\prime}$ for all the zones of the object, with their associated uncertainties, compared with the true values, at a fixed frequency $f_{14}=5.6 \; \mbox{GHz}$. Note that the EM property deviation is important in our example (see figure \ref{fonctions-lambda}). As already mentioned, it is possible to provide some samples of the posterior distribution $p(\mathbf{x}_{14} | \mathbf{y})$ to determine the uncertainty on the estimators. The EM radioelectric properties are correctly inferred all along the ogival object and its 5 material areas. The uncertainty recovers more or less the real profiles. 

\begin{figure}[h]
\centering{\includegraphics[width=1\columnwidth]{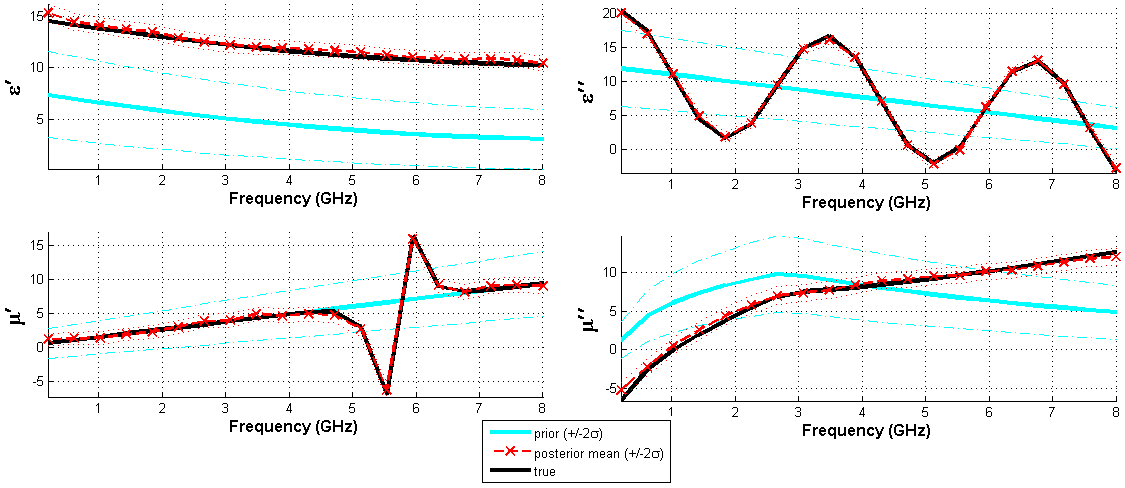}
\caption{\label{ann-fix} Estimated EM properties of the $18^{\mathrm{th}}$ elementary zone }}
\end{figure}

Figure~\ref{ann-fix} presents frequential profiles for a fixed elementary zone (the $18^{\mathrm{th}}$). All the components ($\epsilon^{\prime}, \epsilon^{\prime \prime}, \mu^{\prime}, \mu^{\prime \prime}$) are represented. Each of them is quite accurately estimated. The results are good, even when the perturbations (i.e. the difference between the prior and real profiles) are large and irregular. This robustness is due to the adaptive estimation of $\rho$'s components. Next it is confirmed by several thorough analysis.

\subsection{Performance analysis}   \label{sec-performance}
To extend the results, we propose a statistical performance analysis of the inversion process. It is lead in the same context of section \ref{sec-scenario}. As the developed interacting particle approach is partly stochastic, two different aspects must be studied. Firstly, for a single given data $\mathbf{y}$, the variance of our estimators $\mathbf{\hat{x}}_k$ and $\mathbf{\hat{\Sigma}}_k$, only due to the random feature of the method. Secondly, the average variance of our method for several data $\mathbf{y}^{(i)}$.

\subsubsection{Stochastic variation}
For a given data $\mathbf{y}$, our method mainly provides $2$ sequences of estimators. The posterior mean estimators $(\mathbf{\hat{x}}_1,\ldots,\mathbf{\hat{x}}_{K_f})$, and the posterior covariance matrices estimators $(\mathbf{\hat{\Sigma}}_1,\ldots,\mathbf{\hat{\Sigma}}_{K_f})$. As with all stochastic algorithms, one has to verify that despite random, it always gives the same result, or at least that its own variance is negligible.\\
Let $\mathbf{\hat{x}}$ denote the concatenation of the vectors $\mathbf{\hat{x}}_1,\ldots,\mathbf{\hat{x}}_{K_f}$. Let $\mathbf{\hat{\sigma}}$ denote the concatenation of the estimated marginal uncertainties (square root of the $\mathbf{\hat{\Sigma}}_k$'s diagonal values). Defined in this way, $\mathbf{\hat{x}}$ and $\mathbf{\hat{\sigma}}$ can be considered as $2$ matrices of size $76 \times 20$, and the $2$ main estimators of our method.\\
To quantify the stochastic variance, we simulate an observation data $\mathbf{y}$, and we perform the inversion method $30$ times. At the end, we get $30$ pairs of estimators $\left\{ (\mathbf{\hat{x}}^{(1)} , \mathbf{\hat{\sigma}}^{(1)}) , \ldots , (\mathbf{\hat{x}}^{(30)} , \mathbf{\hat{\sigma}}^{(30)})  \right\}$. For any pair of index $(i,k) \in \{1,\ldots, 76 \} \times \{1,\ldots, 20 \}$, we consider the mean values of the estimators and their RMS (root mean square) values:
$$
\displaystyle{  \mathbf{\bar{\hat{x}}} (i,k) := \frac{1}{30} \sum_{r=1}^{30}  \mathbf{\hat{x}}^{(r)} (i,k)  }  \quad \mbox{and} \quad
\displaystyle{  \mathbf{\bar{\hat{\sigma}}} (i,k) := \frac{1}{30} \sum_{r=1}^{30}  \mathbf{\hat{\sigma}}^{(r)} (i,k)  }
$$
$$
\begin{array}{l}
\displaystyle{ \mbox{RMS} \left( \mathbf{\hat{x}} \right) (i,k) :=  \left( \frac{1}{30} \sum_{r=1}^{30} \left( \mathbf{\hat{x}}^{(r)} (i,k) - \mathbf{\bar{\hat{x}}} (i,k)  \right)^2 \right)^{1/2}  }  \\
\displaystyle{ \mbox{RMS} \left( \mathbf{\hat{\sigma}} \right) (i,k) :=  \left( \frac{1}{30} \sum_{r=1}^{30} \left( \mathbf{\hat{\sigma}}^{(r)} (i,k) - \mathbf{\bar{\hat{\sigma}}} (i,k)  \right)^2 \right)^{1/2}  } 
\end{array}
$$

\paragraph*{}
The numerical results, taken over all the pairs of index $(i,k)$, are summed up in  table \ref{Tab_RMS}.
Two points can be clearly emphasized. First, the standard deviation of the $\mathbf{\hat{x}}^{(r)}$ is very small in an absolute way ($\simeq 10^{-2}$). Moreover, it is negligible compared to the estimated variance of our estimators (at least $1$ decade).
Secondly,  the standard deviation of the $\mathbf{\hat{\sigma}}^{(r)}$ is even smaller ($\simeq 10^{-3}$) and negligible compared to the values of the $\mathbf{\hat{\sigma}}^{(r)}$ themselves (at least $2$ decades). Consequently, there exists a stochastic variance, but it is far negligible compared to the uncertainty inherent to the inverse problem, including measurements.

\begin{table}[ht!]
   \centering
   \begin{tabular}{|c|c|c|c|}
\hline  
mean $\mbox{RMS} \left( \mathbf{\hat{x}} \right)$   &    max $\mbox{RMS} \left( \mathbf{\hat{x}} \right)$   &   mean $\frac{\mbox{\tiny{RMS}} \left( \mathbf{\hat{x}} \right)}{ \mathbf{\bar{\hat{\sigma}}}}$    &   max $\frac{\mbox{\tiny{RMS}} \left( \mathbf{\hat{x}} \right)}{ \mathbf{\bar{\hat{\sigma}}}}$   \\
\hline
$4.16 \; \; 10^{-3}$ & $4.11 \; \; 10^{-2}$ & $1.08 \; \; 10^{-2}$ & $9.87 \; \; 10^{-2}$  \\
\hline
mean $\mbox{RMS} ( \mathbf{\hat{\sigma}} )$   &    max $\mbox{RMS} ( \mathbf{\hat{\sigma}} )$   &   mean $\frac{\mbox{\tiny{RMS}} ( \mathbf{\hat{\sigma}} )}{ \mathbf{\bar{\hat{\sigma}}}}$    &   max $\frac{\mbox{\tiny{RMS}} ( \mathbf{\hat{\sigma}}  )}{ \mathbf{\bar{\hat{\sigma}}}}$   \\
\hline
$1.10 \; \; 10^{-3}$ & $7.56 \; \; 10^{-3}$ & $2.99 \; \; 10^{-3}$ & $1.84 \; \; 10^{-2}$  \\
\hline
\end{tabular}
\caption{\label{Tab_RMS} RMS results of $\mathbf{\hat{x}}$ and $\mathbf{\hat{\sigma}}$}
\end{table}

\subsubsection{Average precision}
The average precision is analyzed on several cases. For this purpose, $30$ independent observation data $\left\{  \mathbf{y}^{(1)}, \ldots , \mathbf{y}^{(30)} \right\}$ are simulated. For each of these observation vectors $\mathbf{y}^{(r)}$, the inversion algorithm computes the pair of estimators $(\mathbf{\hat{x}}^{(r)} , \mathbf{\hat{\sigma}}^{(r)})$. The comparison with the true values of $\mathbf{x}$ is quantified by the following root mean square error (RMSE) :
$$
 \mbox{RMSE} (i,k) :=  \left( \frac{1}{30} \sum_{r=1}^{30} \left( \mathbf{\hat{x}}^{(r)} (i,k) - \mathbf{x}_{\mathrm{true}} (i,k)  \right)^2 \right)^{1/2} 
$$

\begin{figure}[ht!]
\centering{\includegraphics[width=1\columnwidth]{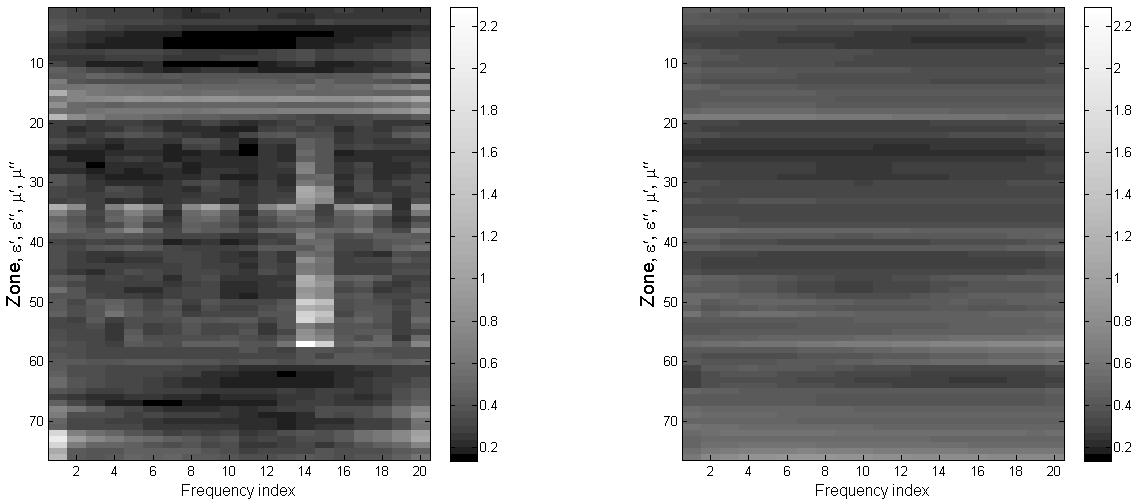}
\caption{\label{RMS-sigma} RMSE (left) and estimated marginal uncertainties (right) }}
\end{figure}

These made errors are shown on figure \ref{RMS-sigma}, where they can be compared to the estimated errors $\mathbf{\bar{\hat{\sigma}}}$. From these results, these conclusions can be drawn. Despite the large amplitude and irregularity of the perturbations, far from the assumed prior model, the estimators $\mathbf{\hat{x}}^{(r)}$ give a good approximation of $\mathbf{x}_{\mathrm{true}} $ (note that the $\mathrm{mean \; RMSE} = 3.68 \; \; 10^{-1}$). Moreover, the RMSE values are comparable to the marginal uncertainties given by $\mathbf{\bar{\hat{\sigma}}}$, which proves that the estimated posterior variances make sense.

\begin{figure}[ht!]
\centering{\includegraphics[width=1\columnwidth]{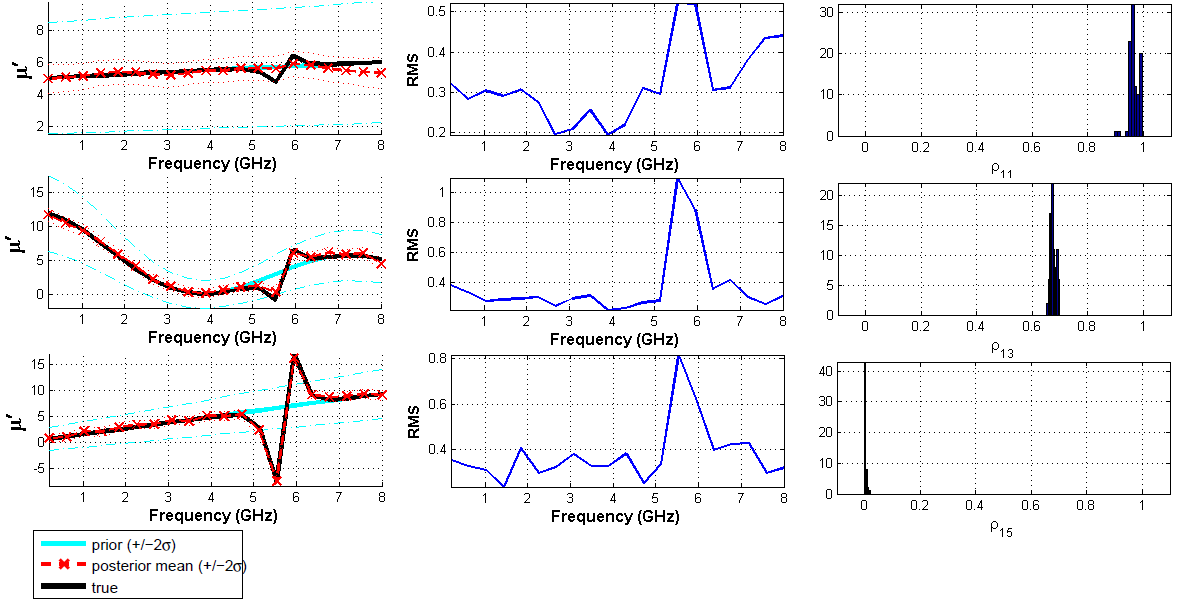}
\caption{\label{freq-RMS-rho} $\mu^{\prime}$ estimators for zones $2$, $9$ and $17$ }}
\end{figure}

In this inversion process, the role of $\rho$'s estimation is very interesting. Roughly speaking, it is as if it can give in advance the shape type of each of the unknown true frequencial profile, by estimating its regularity. On figure \ref{freq-RMS-rho}, we show the results given by $(\mathbf{\hat{x}}^{(1)} , \mathbf{\hat{\sigma}}^{(1)})$ for the zones number $2$, $9$ and $17$ and the permeability $\mu^{\prime}$. On the right part, the histograms represent the posterior distribution of $\rho$.

\noindent As predictable,  the difficulty is increasing from zone $2$ to zone $17$. It is due to the perturbation which is larger and larger, as well as irregular. On the right side of the figure, we show the histograms of all the particles $(\rho^{(1)},\ldots,\rho^{(100)})$ (each particle being represented by its associated component). We clearly see that the more irregular is the true signal, the smaller are the $\rho^{(i)}$, which is quite coherent since $\rho$ quantifies frequential correlation. Meanwhile, we verify in the center of the figure that, in spite of the increasing difficulty, the mean RMSE remains stable. Again, let us stress that the adaptive behavior of $\rho$ estimation is essential to the algorithm robustness.

\subsection{Additional analyses}  \label{sec-variantes}
We propose now to briefly analyze the influence of other parameters, that can come from the context or from the inversion process itself.

\subsubsection{Influence of the Processing Parameters}

The inversion process we described in section \ref{sec-traitement} admits several qualitative and quantitative degrees of freedom, in the SMC step particularly. We propose here our empirical remarks about some of them.
\paragraph{The number of particles $N_p$} Like a classic i.i.d. (independent and identically distributed) sampling method, the SMC algorithm precision is proportional to $N_p^{-1/2}$. However, in our problem, the main objective is not to have a precise estimation of $\eta$, but of $\mathbf{x}$. As the impact of a local variance of $\rho$ on $\mathbf{x}$ is rather small, the crucial point is that the global cloud of particles reaches the correct area in $E$. From this point of view, the important condition is the stability of the Feynman-Kac flow (see \cite{DM-FK}), which ensures that the particles don't get lost in $E$. This is precisely the purpose of the adaptive strategies inside the selection and mutation steps). That's why it seems useless (and time consuming) to use a high number of particles. Note that below $N_p \simeq 40$, the SMC approach may be trapped by some local modes.

\paragraph{The interpolating scheme $\eta_n$} In addition to the annealed probability measure scheme, the hybrid one has been tested. It can assimilate the observations one by one, and update the estimators progressively. Moreover, it manages the computational problems of selectivity that affects the data tempered scheme. The results are good, nearly identical to those obtained with the annealed scheme. And yet, the SMC algorithm lasts around 4 times longer than before. 

\begin{figure}[!h]
\centering{\includegraphics[width=0.8\columnwidth]{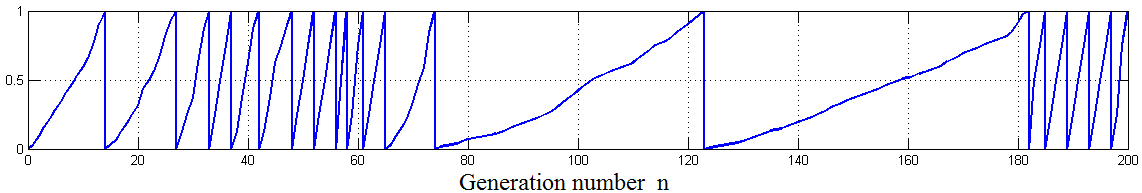}
\caption{\label{alpha-hyb} Annealing parameter $\alpha_i^{(k)}$ }}
\end{figure}

This behavior can be easily interpreted by figure \ref{alpha-hyb}. It appears that many observations do not bring any new information, so that the associated annealing sequence $\alpha_i^{(k)}$ takes value $1$ at once. On the contrary, when a new observation provides information in contradiction with the previous ones, the particles have to migrate from an area of $E$ to another, which takes a longer time (more steps).

\paragraph{The parameter $\rho$} The prior distribution $p(\rho)$ of figure \ref{prior-rho} has a limited impact on the final estimation of $\eta$. Corresponding to a prior knowledge of frequency regularity, it is arbitrary chosen in order to penalize the small values and favor regular profiles. But in practice, this penalization term $p(\rho)$ is less determining than the likelihood one $p(\mathbf{y} | \rho)$. Besides, $\rho$ can be defined $5$-dimensional. In this case, the SMC algorithm performs quicker. However, the underlying hypothesis, i.e. the frequential correlation is the same for $\epsilon^{\prime}$, $\epsilon^{\prime \prime}$, $\mu^{\prime}$, $\mu^{\prime \prime}$, is not necessarily fulfilled in practice.

\subsubsection{Context influence}
As we mentioned, the method is very robust concerning the amplitude and the irregularity of the perturbation (deviation from the reference profiles). 

\paragraph{Measurement noise}  However, it is naturally sensitive to the observation noise magnitude. Its performance degrades when the observation noise is too high. That is clearly a matter of information.  Numerically, it can be explained by considering the accurate approximation given by the surrogate model. Indeed, the $\mathbf{A}_k$ matrices are ill-conditioned. In particular,  $\mu^{\prime}$ and $\epsilon^{\prime \prime}$ components are highly correlated; it is the same for $\mu^{\prime \prime}$ and $\epsilon^{\prime}$. 

\begin{figure}[!h]
\centering{\includegraphics[width=0.8\columnwidth]{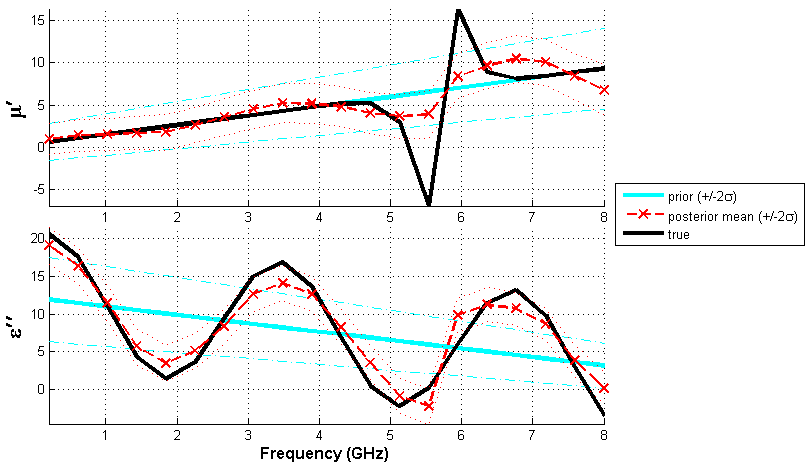}
\caption{\label{Inc10-2} Estimation of $\mu^{\prime}$ and $\epsilon^{\prime \prime}$, $\sigma_{\mathrm{n}} = 10^{-2}$  }}
\end{figure}

In figure~\ref{Inc10-2}, we give the estimations of these $2$ quantities in the case where the amplitude noise $\sigma_{\mathrm{n}} = 10^{-2}$. One can then see that the unknown perturbations of $\mu^{\prime}_{\mathrm{true}}$ and $\epsilon^{\prime \prime}_{\mathrm{true}}$ are correctly detected by the process, but improperly distributed between $\mu^{\prime}$ and $\epsilon^{\prime \prime}$.\\

\paragraph{Problem dimension} Concerning the computation time, it is very sensitive to the dimension of the problem. The reason is simple: each elementary evaluation involves (among others) a Kalman smoother, i.e.  $K_f$ inversions of $4N \times 4N $-sized matrices. Consequently, for a problem of large dimension, it could be appropriate to parallelize the SMC algorithm and distribute the Kalman smoothers on HPC.


\section{Conclusion}
 
An efficient statistical  inference approach has been applied. From  global EM scattering measurements, it manages to estimate local radioelectric properties of materials assembled and placed on the full-scaled object. The inverse problem is solved by combining  intensive computations with high performance computing (HPC), surrogate modeling and advanced sequential Monte Carlo techniques dedicated to frequency dynamic estimation. It takes advantage of the problem structure to achieve a Rao-Blackwellisation strategy of Monte Carlo variance reduction. On top of that, the Bayesian approach quantifies the uncertainties around the estimates.

To tackle higher dimensional problems, it could be interesting to apply close stochastic techniques, such as "interacting Kalman filters", and above all, benefit from the highly parallelization/distribution potential on HPC to tackle 3D geometries and high-dimensional problems.


\section{Annexe: estimation of $\mathbf{x}_k$ and conditioning} \label{SecAnnexe}

For a given  $\mathbf{y}$, it is possible to define judicious estimators of $\mathbf{x}_k$ (for each $k$). Indeed, the first moments $\mathbf{\hat{x}}_k$ and $\mathbf{\hat{\Sigma}}_k)$ can be determined from the theoretical conditional expectation $\mathbf{\bar{x}}_k := \mathbb{E} [\mathbf{x}_k | \mathbf{y}]$ and covariance matrix $\mathbf{\Sigma}_k := \mathbb{V}\mbox{ar} [\mathbf{x}_k | \mathbf{y}]$.\\

For all $\rho \in E$, let set:
$
\mathbf{\hat{x}}_k (\rho) := \mathbb{E}  \left[ \mathbf{x}_k | \rho, \mathbf{y} \right] \quad \mbox{and} \quad
\mathbf{\hat{\Sigma}}_k (\rho) := \mathbb{V}\mbox{ar}  \left[ \mathbf{x}_k | \rho, \mathbf{y} \right]
$, i.e. the main quantities provided by the Kalman smoother. Under this notation, we combine $\hat{\eta} \simeq \eta$ together with the equation (\ref{esp-RB}), and derive a natural choice for the estimator $\mathbf{\hat{x}}_k $:
$$
\begin{array}{ccccccc}
\mathbf{\bar{x}}_k &  =  & \mathbb{E} [\mathbf{x}_k | \mathbf{y}]  & =  &  \mathbb{E}  [ \underbrace{\mathbb{E} (\mathbf{x}_k | \rho,\mathbf{y})}_{\mathbf{\hat{x}}_k (\rho)} | \mathbf{y} ]   &  &  \\
& & &   =   &   \underset{\rho \in E}{\displaystyle{\int}} \mathbf{\hat{x}}_k (\rho)  \eta (d \rho)  & \simeq  &  \underset{\rho \in E}{\displaystyle{\int}} \mathbf{\hat{x}}_k (\rho)  \hat{\eta} (d \rho)   \\
& & &   &   &   =  & \displaystyle{ \underbrace{\frac{1}{N_p} \sum_{i=1}^{N_p}  \mathbf{\hat{x}}_k (\rho^{(i)}) }_{=: \mathbf{\hat{x}}_k}  }
\end{array}
$$
Regarding the covariance estimator $\mathbf{\hat{\Sigma}}_k$, under the same notation and according to (\ref{cov-RB}), we have:
$$
\mathbf{\Sigma}_k  = \underbrace{ \mathbb{E}  \left( \mathbf{\hat{\Sigma}}_k (\rho)  | \mathbf{y} \right) }_{\mathbf{\Sigma}_k^{(1)} }  + \underbrace{ \mathbb{V}\mbox{ar}  \left( \mathbf{\hat{x}}_k (\rho)  | \mathbf{y} \right)  }_{\mathbf{\Sigma}_k^{(2)}} .
$$
We estimate $\mathbf{\Sigma}_k^{(1)}$ and $\mathbf{\Sigma}_k^{(2)}$ separately: 

\begin{enumerate}
\item Evaluation of $\mathbf{\Sigma}_k^{(1)}$
$$
\begin{array}{ccccc}
    \mathbf{\Sigma}_k^{(1)} & = &\mathbb{E}  \left( \mathbf{\hat{\Sigma}}_k (\rho)  | \mathbf{y} \right) \\
     \mathbf{\Sigma}_k^{(1)}  =  &   \underset{\rho \in E}{\displaystyle{\int}}   \mathbf{\hat{\Sigma}}_k (\rho)  \eta (d \rho)  & \simeq & \underset{\rho \in E}{\displaystyle{\int}}   \mathbf{\hat{\Sigma}}_k (\rho)  \hat{\eta} (d \rho)  \\
     &     &  &  = &  \displaystyle{ \underbrace{ \frac{1}{N_p} \sum_{i=1}^{N_p}  \mathbf{\hat{\Sigma}}_k (\rho^{(i)}) }_{=: \mathbf{\hat{\Sigma}}_k^{(1)}}   } 
\end{array}
$$

\item Evaluation of $\mathbf{\Sigma}_k^{(2)}$
$$
\begin{array}{ccccc}
\mathbf{\Sigma}_k^{(2)} & =  &  \mathbb{E}  \left[ \left(\mathbf{\hat{x}}_k (\rho)-\mathbf{\bar{x}}_k  \right) \left(\mathbf{\hat{x}}_k (\rho)-\mathbf{\bar{x}}_k  \right)^{T} | \mathbf{y} \right]  &  &   \\
  &  =  &   \underset{\rho \in E}{\displaystyle{\int}}  \left(\mathbf{\hat{x}}_k (\rho)-\mathbf{\bar{x}}_k  \right) \left(\mathbf{\hat{x}}_k (\rho)-\mathbf{\bar{x}}_k  \right)^{T}  \eta (d \rho)  &  \simeq  &  \underset{\rho \in E}{\displaystyle{\int}}  \left(\mathbf{\hat{x}}_k (\rho)-\mathbf{\hat{x}}_k  \right) \left(\mathbf{\hat{x}}_k (\rho)-\mathbf{\hat{x}}_k  \right)^{T}  \hat{\eta} (d \rho)   \\
  &     &    & =  & \displaystyle{ \underbrace{ \frac{1}{N_p} \sum_{i=1}^{N_p} \left(\mathbf{\hat{x}}_k (\rho^{(i)})-\mathbf{\hat{x}}_k  \right) \left(\mathbf{\hat{x}}_k (\rho^{(i)})-\mathbf{\hat{x}}_k  \right)^{T} }_{=: \mathbf{\hat{\Sigma}}_k^{(2)}  }    }
\end{array}
$$
\end{enumerate}

Finally, the estimator of $\mathbf{\Sigma}_k$ is  given by: $\mathbf{\hat{\Sigma}}_k := \mathbf{\hat{\Sigma}}_k^{(1)} + \mathbf{\hat{\Sigma}}_k^{(2)} $.

\section*{References}
\bibliography{references}
\bibliographystyle{unsrt}
\end{document}